\RequirePackage{lineno}
\documentclass[aps,prl,twocolumn,superscriptaddress]{revtex4-2}%
\usepackage{times,mathptmx}

\usepackage{float}  

\usepackage{graphicx}
\usepackage{graphics}
\usepackage[caption=false]{subfig}
\usepackage{color}
\usepackage[english]{babel}
\usepackage[utf8]{inputenc}
\usepackage{ae}
\usepackage{latexsym}
\usepackage{dcolumn}
\usepackage{bm}
\usepackage[table]{xcolor}
\usepackage{array}
\usepackage{url}
\usepackage{epsf}
\usepackage{epsfig}
\usepackage{pstricks,pst-grad}
\usepackage[pdftex,colorlinks]{hyperref}
\usepackage{amsmath}
\usepackage{amssymb}
\usepackage{ragged2e}
\usepackage{comment}
\usepackage{verbatim}

\begin{document}

\title[]
{Kinetic Pathway and Micromechanics of Vesicle Fusion/Fission}

\author{Luofu Liu}
\affiliation{Department of Chemical and Biomolecular Engineering, University of California Berkeley, Berkeley, California 94720, United States}

\author{Chao Duan}
\affiliation{Department of Chemical and Biomolecular Engineering, University of California Berkeley, Berkeley, California 94720, United States}

\author{Rui Wang}
\email {ruiwang325@berkeley.edu}\affiliation{Department of Chemical and Biomolecular Engineering, University of California Berkeley, Berkeley, California 94720, United States}
\affiliation{Materials Sciences Division, Lawrence Berkeley National Lab, Berkeley, California 94720, United States}


\begin{abstract}
Despite the wide existence of vesicles in living cells as well as their important applications like drug-delivery, the underlying mechanism of vesicle fusion/fission remains under debate. Here, we develop a constrained self-consistent field theory (SCFT) which allows tracking the shape evolution and free energy as a function of center-of-mass separation distance. Fusion and fission are described in a unified framework. Both the kinetic pathway and the mechanical response can be simultaneously captured. By taking vesicles formed by polyelectrolytes as a model system, we predict discontinuous transitions between the three morphologies: parent vesicle with a single cavity, hemifission/hemifusion and two separated child vesicles, as a result of breaking topological isomorphism. With the increase of inter-vesicle repulsion, we observe a great reduction of the cleavage energy, indicating that vesicle fission can be achieved without hemifission, in good agreement with simulation. The force-extension relationship elucidates typical plasticity for separating two vesicles. The super extensibility in the mechanical response of vesicle is in stark contrast to soft particles with other morphologies such as cylinder and sphere.
\end{abstract}
\maketitle

Vesicle is one of the most important structures formed via self-assembly \cite{Mai2012,Karayianni2021,Blanazs2009,Discher2002}. 
In biology, vesicles widely exist within and outside cells, which constitutes the basic structure of cells and organelles, governing a variety of functions of human body. Owing to the unique hollow
morphology, vesicles synthesized by amphiphilic molecules also have a wealth of applications such as encapsulators, micro/nanoreactors, and drug/gene delivery cargos \cite{Jiang2022,Wang2018,Lv2022,Axthelm2008,Palivan2012}. Fusion and fission are two vital processes for the functionality of vesicles. Many biological processes including cell division, endocytosis and exocytosis are intimately related to vesicle fusion/fission \cite{gelbart2012,johnson2002,Jahn2003,Jahn2012}. Fusion/fission also have crucial impacts on the encapsulating and releasing properties of vesicles as nano/biomadical materials, such as the uptake rate of lipsome in drug delivery \cite{Düzgüneş1999,Poon2020,Shen2019,Ho2021}.

Despite the ubiquity in nature and the wide applications, understanding the mechanism of vesicle fusion/fission remains a great challenge. Although similar intermediate structures were observed in transmission electron microscopy (TEM) images for both fission and fusion \cite{Luo2001}, it is generally accepted that these two processes go through different kinetic pathways \cite{Smeijers2006, Markvoort2007}.
The pioneering picture to describe the vesicle fusion is the ``stalk" model  \cite{Markin1984,Chernomordik1987,Chernomordik1995}, which suggests that the fusion is initiated by forming a stalk, followed by a radial expansion of the stalk to the hemifusion diaphragm, and ends up with the pore-opening. However, there is discrepancy between the assumptions in the stalk model and the observations in simulations \cite{Grafmüller2007,Gao2008,Li2009,Risselada2012,Marrink2003,Shillcock2005}. Whether the stalk undergoes a substantial expansion and in which direction the stalk grows is still under debate \cite{Siegel1999,Siegel1993,Noguchi2001}. On the other hand, the most classic model to describe vesicle fission is the budding-fission mechanism \cite{Matsuoka1998,Spang1998,Takei1998}. It is featured by the formation of the hemifission where the inner layer of the parent vesicle merges into a neck followed by cleavage into two child vesicles \cite{Kozlovsky2003,Shemesh2003,Zimmerberg2005,Atilgan2007,Campelo2008,Kozlov2010,Kozlov2022}. Nevertheless, recent simulations showed that hemifission is not necessary: the parent vesicle can directly split into two child vesicles without neck formation \cite{Li2009,Yamamoto2003}. Furthermore, micromechanics of the vesicle bilayers in response to the applied stress also arouses great interest. Using coarse-grained simulation, Park et al. observed a tubulation with extremely large extensibility when pulling a vesicle from a membrane \cite{Park2019}.

Many theoretical efforts have been made to explain the mechanism of vesicle fusion/fission. Most descriptions are based on elastic models, which capture the shape evolution and the corresponding energy of the intermediate states \cite{Siegel1999,Siegel1993,Kozlovsky2003}. It is difficult for this phenomenological treatment to include molecular structure and interactions on the pathway. For another approach, Schick et al. applied the self-consistent field theory (SCFT) to study the bilayers made of diblock copolymers, which allows to capture the coupling between the molecular conformation and the shape change of the bilayer \cite{Katsov2004,Katsov2006,Lee2007,Lee2008}. Using the radius of stalk as the reaction coordinate, the stability of intermediates was analyzed. Being focused on a small portion near the stalk, their method however fails to capture the shape change of the entire vesicle. 

In this work, we develop a constrained SCFT to study the fusion/fission of two vesicles formed by polyelectrolytes (PE). The molecular structure and interactions are systematically included. By tracking shape evolution and the corresponding free energy as a function of center-of-mass (c.m.) separation distance, we are able to describe the fusion and fission processes in a unified framework. The kinetic pathway and the mechanical response can be obtained simultaneously. We predict that the transitions between different morphologies are discontinuous due to the breaking of topological isomorphism. We observe super extensibility when separating two vesicles, which resembles typical plastic materials. Our predictions are in good agreement with simulation results. 

\begin{figure}[h] 
\centering
\includegraphics[width=0.45\textwidth]{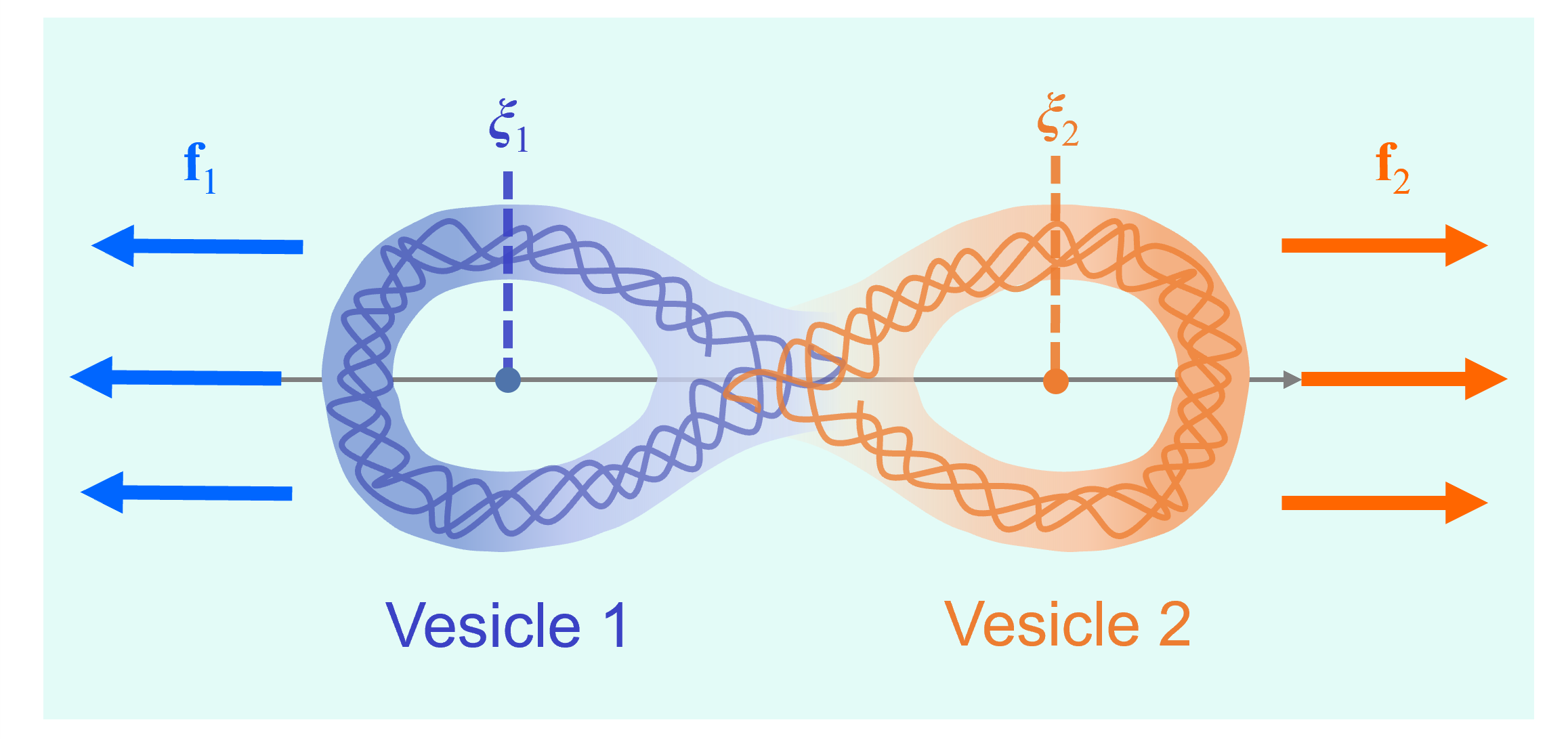}
\caption{Schematic of fusion/fission of two vesicles formed by polyelectrolytes. The centers of mass of the two vesicles are fixed at $\boldsymbol{\xi}_1$ and $\boldsymbol{\xi}_2$ with a separation distance $L = |\boldsymbol{\xi}_2 - \boldsymbol{\xi}_1|$. External forces ($\mathbf{f}_1 = -\mathbf{f}_2$) are applied to maintain the separation.}
\label{figure:schematic}
\end{figure}

Previous work showed that stable vesicle can be formed by a single uniformly-charged polyelectrolyte (PE) \cite{Duan2022}. Here, we take this vesicle as a model to study the fusion/fission process. The system we consider is a semicanonical ensemble consisting of two PEs and $n_{\rm s}$ solvent molecules in the presence of $n_{+}$ cations and $n_{-}$ anions. The number of PE is fixed while the solvent and mobile ions are connected with a bulk salt solution of an ion concentration $c_{\pm}^{\rm b}$ that maintains the chemical potentials of the solvent $\mu_{\rm s}$ and ions $\mu_{\pm}$. PEs are assumed to be Gaussian chains of $N$ Kuhn segments with Kuhn length $b$. The backbone charge density is $\alpha$. Mobile ions are taken as point charges with valency $z_{\pm}$. The two vesicles are identical but distinguishable, denoted as Vesicle 1 and Vesicle 2 shown in Fig. \ref{figure:schematic}. The c.m. of the two vesicles are fixed at position $\boldsymbol{\xi}_{1}$ and $\boldsymbol{\xi}_{2}$, respectively. 

The partition function of the system can be written as: 
\begin{equation} \label{eq:partition_fucntion}
\begin{aligned}
    \boldsymbol{\Xi} &= \frac{1}{v_{\rm p}^{2N}} \sum_{n_{\gamma}=0}^\infty \prod_{\gamma} \frac{e^{\mu_{\gamma} n_{\gamma}}}{n_{\gamma}!v_{\gamma}^{n_{\gamma}}}\prod_{j=1}^2\int {\rm \hat{D}}\{\mathbf{R}_j\} \prod_{\kappa = 1}^{n_{\gamma}} \int{\rm d}\mathbf{r}_{\gamma,\kappa}  \\
    &\exp(-\mathcal{H}) \prod_{\mathbf{r}}\delta \left[\hat{\phi}_{\rm p}(\mathbf{r})+\hat{\phi}_{\rm s}(\mathbf{r})-1 \right]  \\
    &\times \delta\left[\frac{1}{N}\int_{0}^{N}{\rm d}s\mathbf{R}_1(s)-\boldsymbol{\xi}_1\right]\delta\left[\frac{1}{N}\int_{0}^{N}{\rm d}s\mathbf{R}_2(s)-\boldsymbol{\xi}_2\right] \\
\end{aligned}
\end{equation}
where $\gamma = {\rm s}, \pm$ stands for solvents, cations and anions, respectively. $\int \hat{D}\{\mathbf{R}_j\}$ denotes the integration over the Gaussian-weighted chain configurations. $v_{\rm p}$ and $v_\gamma$ are the volumes of the chain segments and small molecules. We assume $v_{\rm p} = v_{\rm s} = v$. $\hat{\phi}_{\rm p} = \hat{\phi}_{\rm p1} + \hat{\phi}_{\rm p2}$ is the total instantaneous volume fractions of PE, and $\hat{\phi}_{\rm s}$ is that of solvent. The $\delta$ functional in the second line of Eq. \ref{eq:partition_fucntion} accounts for the incompressibility. The Hamiltonian $\mathcal{H}$ is given by
\begin{equation} \label{eq:Hamiltonian}
\mathcal{H} =  \frac{\chi}{v}\int{\rm d}{\mathbf{r}}\hat{\phi}_{\rm p}(\mathbf{r})\hat{\phi}_{\rm s}(\mathbf{r}) + \frac{1}{2}\int {\rm d}{\mathbf{r}}\int {\rm d}{\mathbf{r^{'}}} \hat{\rho}(\mathbf{r}) C(\mathbf{r}, \mathbf{r^{'}})\hat{\rho}(\mathbf{r}^{'})
\end{equation}
where the two contributions come from the short-range polymer-solvent interaction characterized by the Flory-Huggins $\chi$ parameter and the long-range electrostatic interaction between all charged species. $\hat{\rho}(\mathbf{r}) = z_{+}\hat{c}_{+}(\mathbf{r}) - z_{-}\hat{c}_{-}(\mathbf{r}) - \alpha \hat{\phi}_{\rm_p}(\mathbf{r})/v$ is the instantaneous local charge density, with $\hat{c}_{\pm}(\mathbf{r})$ the instantaneous ion concentration. $C(\mathbf{r}, \mathbf{r^{'}})$ is the Coulomb operator, satisfying $-\nabla \cdot[\epsilon(\mathbf{r})\nabla C(\mathbf{r},\mathbf{r^{'}})] = \delta(\mathbf{r} -\mathbf{r^{'}})$.
$\epsilon(\mathbf{r}) = \epsilon_0\epsilon_{\rm r}(\mathbf{r})/\beta e^2$ is the rescaled permittivity, where  $\epsilon_0$ is the vacuum permitivity, $e$ is the elementary charge and $\epsilon_{\rm r}$ is local dielectric constant \cite{Issei2012,wang2011jcp,sing2014,Kevin2018}. The two $\delta$ functions in the third line of Eq. \ref{eq:partition_fucntion} are introduced to enforce the c.m. of the Vesicle $j$ at $\xi_j$ \cite{Liu2022, grosberg1992}.
We follow the standard SCFT techniques \cite{fredrickson2006equilibrium} (see the detailed derivation in Sec. I of the Supplementary Material). The identity transformation of the $\delta$ functions that account for the constraint of c.m. generates two force fields $\mathbf{f}_j$, conjugating to the deviation of c.m. The force fields counteract the inter-vesicle interaction, which guarantees the sampling of all configurations along the kinetic pathway. The resulting self-consistent equations for PE density $\phi_{{\rm p}j}(\mathbf{r})$, electrostatic potential $\psi(\mathbf{r})$, and conjugate fields $w_{\rm p,s}(\mathbf{r})$ are:
\begin{subequations}\label{scfs}
\begin{align}
&w_{\rm p}(\mathbf{r}) - w_{\rm s}(\mathbf{r})=\chi\left[1-2\phi_{\rm p}(\mathbf{r})\right]-v\frac{\partial \epsilon}{\partial \phi_{\rm p}}\frac{|\nabla \psi|^2}{2}-\alpha \psi(\mathbf{r}) \label{scf_a}\\
&\phi_{{\rm p}j}(\mathbf{r}) = \frac{1}{Q_{{\rm p}j}}\int_0^{N}{\rm d}s q_j(\mathbf{r},s)q_j(\mathbf{r},N-s)\label{scf_b} \\
&1-\phi_{\rm p}(\mathbf{r}) = \exp\left[ \mu_{\rm s}-w_{\rm s}(\mathbf{r})\right] \label{scf_c} \\
& -\nabla \cdot \left[\epsilon(\mathbf{r})\nabla \psi(\mathbf{r})\right]=z_{+}c_{+}(\mathbf{r})-z_{-}c_{-}(\mathbf{r})-\frac{\alpha}{v}\phi_{\rm p}(\mathbf{r}) \label{scf_d} \\
& {\int {\rm d}\mathbf{r}\phi_{{\rm p}j}(\mathbf{r})\cdot(\mathbf{r}-\boldsymbol{\xi}_j)} = \mathbf{0} \label{scf_e}
\end{align}
\end{subequations}
$c_{\pm}(\mathbf{r}) = \lambda_{\pm}\exp(\mp z_{\pm}\psi(\mathbf{r}))$ is the ion concentration, where $\lambda_{\pm} = \exp(\mu_{\pm})/v_{\pm}$ is the fugacity of ions determined by $c_{\pm}^{\rm b}$. $Q_{{\rm p}j} = v^{-1}\int {\rm d}{\mathbf{r}}q_j(\mathbf{r},N)$ is the single chain partition function of Vesicle $j$ in the fields $w_{\rm p}(\mathbf{r})$ and $\mathbf{f}_{j}$. $q_j$ is the propagator satisfying the modified diffusion equation:
\begin{equation}
    \left\{ \frac{\partial }{\partial s} - \frac{b^2}{6}\nabla^2 + \left[ w_{\rm p}(\mathbf{r})-\frac{\mathbf{f}_j}{N}\cdot(\mathbf{r}-\boldsymbol{\xi}_j) \right]  \right\}q_j(\mathbf{r},s) = 0
    \label{eq:propagator}
\end{equation}
It can be seen that 
the external field is shifted by the force ${\mathbf f}_j$ as a result of constraining c.m., such that the shape of vesicles will be deformed from the force-free system. The free energy $F$ of the two vesicles is:
\begin{equation}\label{eq:free_energy}
\begin{aligned}
    F_{\rm} &= -e^{ \mu_{\rm s}}Q_{\rm s} - \ln Q_{{\rm p}1} - \ln Q_{{\rm p}2} \\
    &+ \frac{1}{v}\int {\rm d}{\mathbf{r}}\left[\chi\phi_{\rm p}\left(1-\phi_{\rm p}\right)
    -w_{\rm p}\phi_{\rm p}-w_{\rm s}\left(1- \phi_{\rm p} \right)\right]\\
    &-\int {\rm d} {\mathbf{r}}\left[c_{+} + c_{-} + \frac{1}{2}\epsilon(\nabla \psi)^2 + \frac{\alpha}{v}\phi_{\rm p}\psi \right]
\end{aligned}
\end{equation}
where $Q_{\rm s} = v^{-1}\int {\rm d}\mathbf{r}\exp[-w_{\rm s}(\mathbf{r})]$ is the solvent partition function. The vesicle density profile and the free energy as a function of separation distance $L$ can be obtained by solving Eqs. \ref{scfs} and \ref{eq:propagator} iteratively. For the two vesicle system, ${\mathbf{f}_1} = -\mathbf{f}_2$; thus only one force is needed in the iteration. $\mathbf{f}_j$ is determined by using Eq. \ref{scf_e} as a 
Lagrangian multiplier. The numerical details are provided in Sec. II of the Supplementary Material.   

In the current work, we focus on the dependence of kinetic pathway and micromechanics of fusion/fission on the essential topological feature of the vesicle. We take the vesicle formed by PE as a model system. The fundamental physics obtained here can be applied to other types of vesicles, such as those formed by sufactants, lipids and block copolymers. Moreover, the constrained SCFT developed here can be easily generalized to study the interactions, kinetic process and micromechanics of various soft matter aggregates, such as polymer chains, coacervates, intrinsically disordered proteins, micelles, nano/micro gels and polymer-grafted nanoparticles.

\begin{figure}[t] 
\centering
\includegraphics[width=0.45\textwidth]{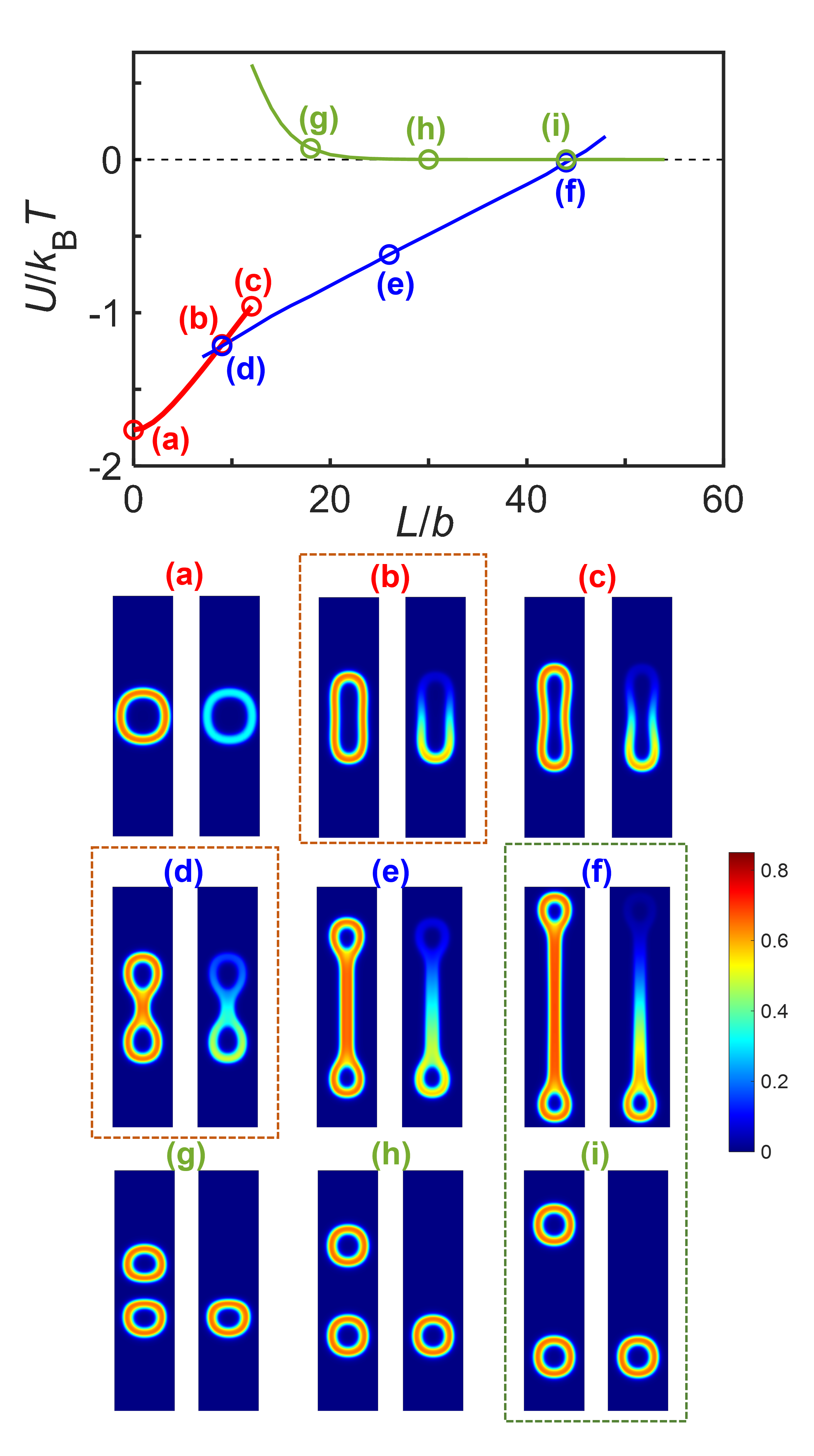}
\caption{Kinetic pathway of vesicle fission/fusion. Potential of mean force $U$ along c.m. separation distance $L$ is shown on the top. The three types of morphology: parent vesicle with a single cavity, hemifission/hemifusion and two separated child vesicles are represented by red, blue and green curves respectively. The bottom panel shows 2D visualization of the density profile for the representative states. The colorbar denotes the PE volume fraction. The dashed frames of the same color pair the coexistent states in the kinetic pathway. For each state, the left figure plots the entire volume fraction of the two vesicles, whereas the right figure plots that of one constituting vesicle. The ranges for the $r$- and $z$-axis are $r/b\in [-10, 10]$ and $z/b \in [-40, 40]$. $N=200$, $b=1 \rm{nm}$, $\chi=1.25$, $\alpha=0.72$, $\epsilon_{\rm r}=80$, $z_+ = z_- = 1$ and $c_{\rm b}=0.10{\rm M}$.} 
\label{figure:PMF}
\end{figure}

Figure \ref{figure:PMF} shows the free energy in terms of potential of mean force $U(L)$ between two vesicles and the representative morphologies. $U(L) = F(L) - F(\infty)$, is the free energy at a given $L$ excessive to the reference state at $L=\infty$. Going along the fission direction, three different features of morphology can be observed: (1) parent vesicle with a single cavity, (2) hemifission/hemifusion characterized by a neck connecting two cavities, and (3) two separated ``child" vesicles. From State (a) to State (c), the parent vesicle is elongated; each of the two constituting vesicles is polarized to one end. As $L$ reaches a critical value (State (b)), two opposing inner layers merge into a neck (State (d)), signifying hemifission/hemifusion. Neck growth spans a long range of $L$ (State (d) to State (f)), where the neck width remains almost constant and the cavity volume shrinks. Lastly, the fission process ends up with the fracture of the neck,  by forming two separated child vesicles (State (i)). The morphological evolution predicted by our theory is in good agreement with the intermediate structures observed in TEM images \cite{Luo2001}.

The energy landscape reveals the characteristics of the morphological transition. Figure \ref{figure:PMF} shows that both the transition from the elongated parent vesicle to hemifission/hemifusion and the subsequent transition to two separated child vesicles are discontinuous, demonstrated by the crossing of their corresponding free energy curves. Unlike the common first-order phase transitions as a result of symmetry breaking, the discontinuous transition here is driven by the change of topological isomorphism belonging to different morphologies. The morphological transitions hence have to involve a nucleation process with an activation energy \cite{Xu2014,Bottacchiari2022}. It should be noted that this nucleation cannot be captured in our current approach using c.m. separation distance as the reaction coordinate. It can be accounted for by more sophisticated transition-path theory such as string method \cite{Weinan2010,Xu2014,Bottacchiari2022}.

Another important implication of the discontinuous transition is the existence of metastable regions beyond the exact transition point, such that the shape evolution in fusion/fission does not need to exactly follow the minimum free energy curve. It provides a possibility that fusion and fission undergo different pathways in the real process, if each includes different portions of the metastable regions. Therefore, fission and fusion are not reverse to each other and have to be described differently. Taking the fusion as an example, the transition from two separated child vesicles to hemifusion is not likely to occur at State (i). The nucleation barrier for such transition is expected to be high, because the two vesicles cannot tell each other restricted by the short-range hydrophobic attraction. To reduce the nucleation barrier, the fusion would rather proceed along the metastable path (from State (i) to (g)) until the two vesicles get sufficiently approached. As a result, the long neck predicted by our theory in State (f) and also observed in the simulation of fission \cite{Park2019} would not appear in fusion. 

For another example, the metastable state of the two separated child vesicles can transit to the morphology of either the hemifusion or the elongated parent vesicle. This is confirmed by the existence of all three morphologies as the equilibrium structure (either stable or metastable) at the same $L=12b$. If the former transition occurs, there is a window for stalk growth prior to pore-opening, in line with the original stalk model \cite{Markin1984,Chernomordik1987,Chernomordik1995}. If the latter transition occurs, pore will open without substantial stalk growth, consistent with modified stalk model \cite{Siegel1993,Siegel1999}. Which of the two transitions is preferred depends on their corresponding nucleation barriers. There is a long-time debate on the intrinsic mechanism of vesicle fusion. Our theory provides a unified description of different mechanisms.

\begin{figure}[t] 
\centering
\includegraphics[width=0.4\textwidth]{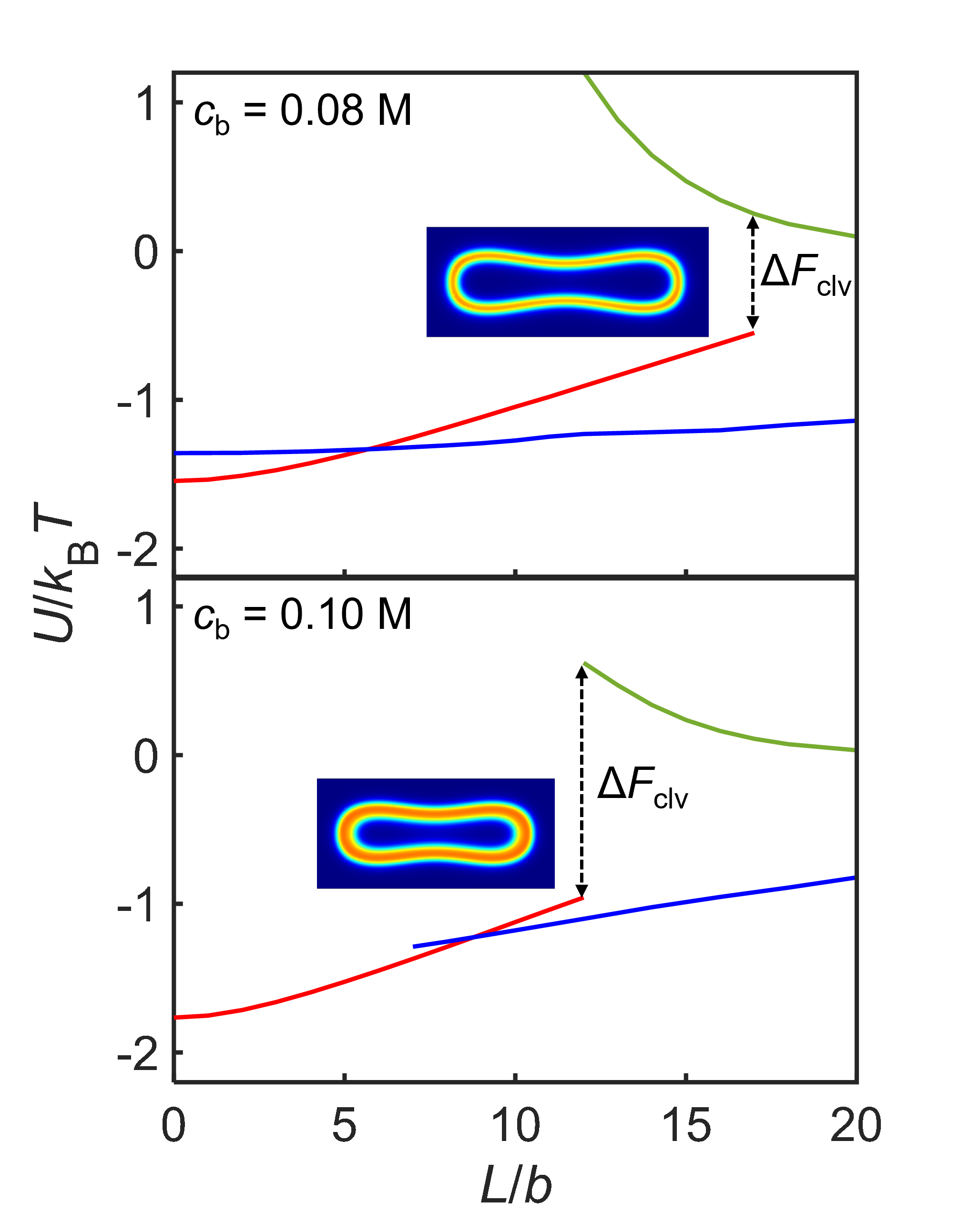}
\caption{Effect of electrostatic repulsion on the fission mechanism. $U$ is plotted against $L$ for $c_{\rm b}=0.08 {\rm M}$, where the case of weak repulsion ($c_{\rm b}=0.10 {\rm M}$) is shown in comparison. The dash line indicates the energy gap $\Delta F_{\rm clv}$ between the elongated parent vesicle and the two child vesicles at the fracture point, where the insets visualize the corresponding morphology of the parent vesicle.}
\label{figure:effect_of_salt}
\end{figure}

The classical pathway of fission observed in many experiments and simulations follows the budding-fission mechanism involving a necking process. However, this is challenged by the findings that the intermediate hemifission step can be bypassed: the parent vesicle is directly cleaved into two child vesicles \cite{Li2009,Yamamoto2003}. This alternative mechanism can also be captured by our theory. Figure \ref{figure:effect_of_salt} shows the expansion of the metastable region of the parent vesicle as the electrostatic repulsion increases. The parent vesicle with a single cavity can be elongated to a much larger $L$ before it gets unstable. The vesicle shell becomes extremely thin with a much lower PE density inside as shown in the inset of Fig. \ref{figure:effect_of_salt}, implying a smaller elastic modulus upon deformation and hence a lower resistance to fracture. The cleavage energy $\Delta F_{\rm clv}$, defined by the energy gap between the two states at the fracture point, is also greatly reduced. $\Delta F_{\rm clv}$ is much lower than $k_{\rm B}T$ for the case of very strong repulsion. Therefore, the thin shell of the elongated parent vesicle is easy to be fractured by thermal fluctuation, which induces a direct transition to two child vesicles. Both Li et al. \cite{Li2009} and Yamamoto et al. \cite{Yamamoto2003} observed that fission without necking occurs in the presence of strong repulsive interactions, in good agreement with our theoretical prediction.

\begin{figure}[t] 
\centering
\includegraphics[width=0.45\textwidth]{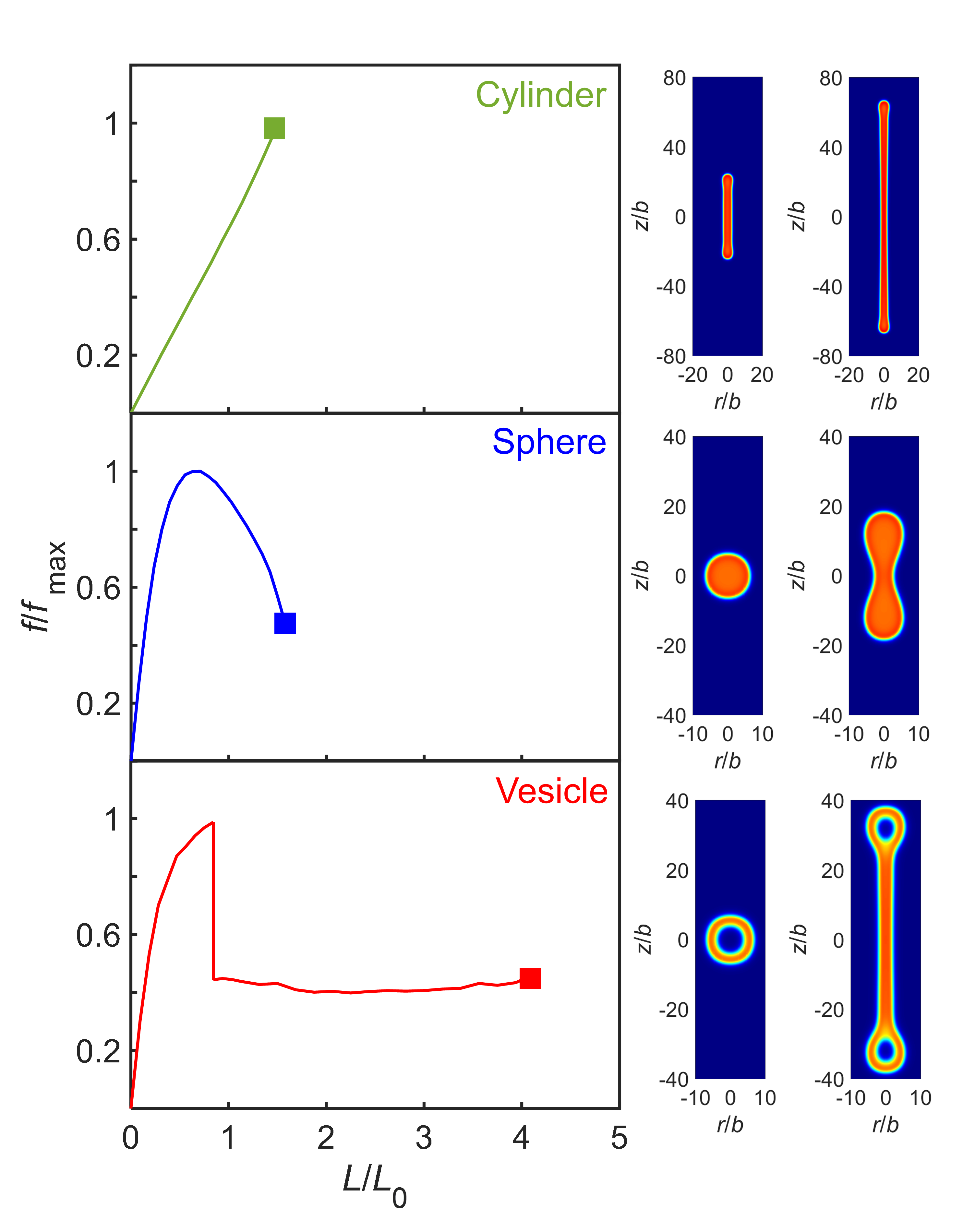}
\caption{Micromechanics in vesicle fission compared to the cases of cylinder and sphere. The left panel plots the rescaled tensile forces $f/f_{\rm max}$ for separating two soft particles against the normalized c.m. separation distance $L/L_0$. $f_{\rm max}$ are the maximum values of the force. $L_0$ are the characteristic lengths of a single force-free particle, i.e. the diameter of vesicle and sphere, and the axial length of cylinder. The symbols denote the fracture points. The right panel visualizes super extensibility of vesicle. The morphologies of the two associated particles at the fracture point are shown in right figures in comparison with those of a single force-free particle (left figures). $\alpha=0.72$, $c_{\rm b}=0.10{\rm M}$ for vesicle, $\alpha=0.50$, $c_{\rm b}=0.05{\rm M}$ for cylinder, and $\alpha=0.72$, $c_{\rm b}=0.15{\rm M}$ for sphere.}
\label{figure:Micromechanics}
\end{figure}

Our constrained SCFT simultaneously captures the kinetic pathway and the microscopic mechanical response. Figure \ref{figure:Micromechanics} quantifies the tensile force $f$ needed to separate two associated vesicles as a function of $L$, which presents a microscopic analog to the stress-strain characterization in material mechanics. For comparison, we also show the behavior of 
another two types of soft particles, cylinders and spheres, which are formed by the same type of PE as vesicle but with different $\alpha$ and $c_{\rm b}$ \cite{Duan2022}. The three soft particles show exact different mechanical response to the applied force. Pure elasticity is observed for cylinders. Force increases almost linearly with the elongation before an abrupt breaking, resembling pure elastic materials such as brittle ceramics. Yielding feature is observed for spheres: $f$ increases for small deformation, reaches maximum and drops until fracture. This resembles ductile materials such as common metals. In stark contrast, vesicle exhibits typical plastic behavior. A long plateau with almost constant force appears after yielding, which resembles plastic materials such as crystalline polymers \cite{ward2012mechanical}. Furthermore, Fig. \ref{figure:Micromechanics} also elucidates the super extensibility of vesicle in response to the tensile force compared to cylinder and sphere. The two associated vesicles can be elongated to more than 4 times the diameter of the single vesicle before fracture. The super extensibility predicted here is in good agreement with recent simulation by Park et al. when pulling a vesicle from a bilayer membrane \cite{Park2019}. It is interesting to note that the mechanics for different macroscopic materials can also be applied to soft nanoparticles at microscopic level.

The super extensibility exhibited in vesicle fission can be understood as follows. Geometrically, the increase of surface area during the neck growth is negligible for vesicle compared to cylinder and sphere (see Sec. III in the Supplementary Material). The reduction of the area from both the outer and inner surfaces of the vesicle shells compensates the increase of the area from neck, resulting in little penalty in surface energy during neck growth. Mechanically, the very low surface tension of vesicle imposes small resistance to the expansion of surface area when pulling two vesicles apart \cite{gelbart2012,brochard1976}. Materials continuously flow from cavity shells to neck (manifested as ``in-plane flow" \cite{Arroyo2009}), resembling the alignment of chain segments in crystalline polymers under tension. Thermodynamically, the stable regimes of vesicle and cylinder are adjacent in phase diagram, which holds for all kinds of vesicles regardless of the constituting molecules, e.g. surfactants \cite{steed2012supramolecular}, block copolymers \cite{Discher2002}, PEs \cite{Duan2022}. The transition from vesicle to cylinder involves the breaking of both spherical symmetry and topological isomorphism, which necessities a two-step nucleation with the existence of an intermediate state \cite{Duan2023}. Previous work found a metastable intermediate named "vesicle-necklace" near the phase boundary \cite{Duan2022}. The highly extensible necking structure formed during vesicle fission is essentially a vesicle-necklace and thus can be stabilized against fracture. 

In this Letter, we develop a constrained SCFT to study vesicle fusion/fission. The theory systematically includes molecular structure and interactions. By tracking the shape evolution and the corresponding free energy as a function of c.m. separation distance, our theory simultaneously captures the kinetic pathway and the mechanical response. We observe three features of morphology along the pathway: parent vesicle with a single cavity, hemifusion/hemifission, and two separated child vesicles. The transitions between these morphologies are discontinuous as a result of breaking topological isomorphism. The existence of metastable regions implies that fusion and fission can undergo different kinetic pathways in the real process. Our theory provides a unified description of both the original and modified stalk model in vesicle fusion. With the increase of inter-vesicle repulsion, we observe a significant reduction of the cleavage energy, indicating that vesicle fission can be achieved without hemifission. The force-extension relationship for separating two vesicles shows a long plateau, a typical feature of plastic materials. This is in stark contrast to other soft particles, where cylinder shows pure elasticity and sphere resembles ductile metals. Our theoretical predictions on morphology, kinetics and mechanics are in good agreement with the observations in experiments and simulations. Although the PE vesicle is taken as a model system, we believe that the fundamental physics based on the intrinsic topological features is universal for vesicles formed by other constituting molecules, like surfactants, lipids, block copolymers, etc. Furthermore, our constrained SCFT can be generalized to study the interactions, kinetic process and micromechanics of various soft matter systems such as polymer chains, coacervates, intrinsically disordered proteins, micelles, nano/micro- gels and polymer-grafted nanoparticles.

Acknowledgment is made to the donors of the American Chemical Society Petroleum Research Fund for partial support of this research. This research used the computational resources provided by the Kenneth S. Pitzer Center for Theoretical Chemistry.

\bibliographystyle{apsrev4-2}
\bibliography{manuscript}

\begin{thebibliography}{64}%
\makeatletter
\providecommand \@ifxundefined [1]{%
 \@ifx{#1\undefined}
}%
\providecommand \@ifnum [1]{%
 \ifnum #1\expandafter \@firstoftwo
 \else \expandafter \@secondoftwo
 \fi
}%
\providecommand \@ifx [1]{%
 \ifx #1\expandafter \@firstoftwo
 \else \expandafter \@secondoftwo
 \fi
}%
\providecommand \natexlab [1]{#1}%
\providecommand \enquote  [1]{``#1''}%
\providecommand \bibnamefont  [1]{#1}%
\providecommand \bibfnamefont [1]{#1}%
\providecommand \citenamefont [1]{#1}%
\providecommand \href@noop [0]{\@secondoftwo}%
\providecommand \href [0]{\begingroup \@sanitize@url \@href}%
\providecommand \@href[1]{\@@startlink{#1}\@@href}%
\providecommand \@@href[1]{\endgroup#1\@@endlink}%
\providecommand \@sanitize@url [0]{\catcode `\\12\catcode `\$12\catcode
  `\&12\catcode `\#12\catcode `\^12\catcode `\_12\catcode `\%12\relax}%
\providecommand \@@startlink[1]{}%
\providecommand \@@endlink[0]{}%
\providecommand \url  [0]{\begingroup\@sanitize@url \@url }%
\providecommand \@url [1]{\endgroup\@href {#1}{\urlprefix }}%
\providecommand \urlprefix  [0]{URL }%
\providecommand \Eprint [0]{\href }%
\providecommand \doibase [0]{https://doi.org/}%
\providecommand \selectlanguage [0]{\@gobble}%
\providecommand \bibinfo  [0]{\@secondoftwo}%
\providecommand \bibfield  [0]{\@secondoftwo}%
\providecommand \translation [1]{[#1]}%
\providecommand \BibitemOpen [0]{}%
\providecommand \bibitemStop [0]{}%
\providecommand \bibitemNoStop [0]{.\EOS\space}%
\providecommand \EOS [0]{\spacefactor3000\relax}%
\providecommand \BibitemShut  [1]{\csname bibitem#1\endcsname}%
\let\auto@bib@innerbib\@empty
\bibitem [{\citenamefont {Mai}\ and\ \citenamefont
  {Eisenberg}(2012)}]{Mai2012}%
  \BibitemOpen
  \bibfield  {author} {\bibinfo {author} {\bibfnamefont {Y.}~\bibnamefont
  {Mai}}\ and\ \bibinfo {author} {\bibfnamefont {A.}~\bibnamefont
  {Eisenberg}},\ }\href
  {https://pubs.rsc.org/en/content/articlelanding/2012/cs/c2cs35115c}
  {\bibfield  {journal} {\bibinfo  {journal} {Chem. Soc. Rev.}\ }\textbf
  {\bibinfo {volume} {41}},\ \bibinfo {pages} {5969} (\bibinfo {year}
  {2012})}\BibitemShut {NoStop}%
\bibitem [{\citenamefont {Karayianni}\ and\ \citenamefont
  {Pispas}(2021)}]{Karayianni2021}%
  \BibitemOpen
  \bibfield  {author} {\bibinfo {author} {\bibfnamefont {M.}~\bibnamefont
  {Karayianni}}\ and\ \bibinfo {author} {\bibfnamefont {S.}~\bibnamefont
  {Pispas}},\ }\href
  {https://onlinelibrary.wiley.com/doi/full/10.1002/pol.20210430} {\bibfield
  {journal} {\bibinfo  {journal} {J. Polym. Sci.}\ }\textbf {\bibinfo {volume}
  {59}},\ \bibinfo {pages} {1874} (\bibinfo {year} {2021})}\BibitemShut
  {NoStop}%
\bibitem [{\citenamefont {Blanazs}\ \emph {et~al.}(2009)\citenamefont
  {Blanazs}, \citenamefont {Armes},\ and\ \citenamefont {Ryan}}]{Blanazs2009}%
  \BibitemOpen
  \bibfield  {author} {\bibinfo {author} {\bibfnamefont {A.}~\bibnamefont
  {Blanazs}}, \bibinfo {author} {\bibfnamefont {S.~P.}\ \bibnamefont {Armes}},\
  and\ \bibinfo {author} {\bibfnamefont {A.~J.}\ \bibnamefont {Ryan}},\ }\href
  {https://onlinelibrary.wiley.com/doi/full/10.1002/marc.200800713} {\bibfield
  {journal} {\bibinfo  {journal} {Macromol. Rapid Commun.}\ }\textbf {\bibinfo
  {volume} {30}},\ \bibinfo {pages} {267} (\bibinfo {year} {2009})}\BibitemShut
  {NoStop}%
\bibitem [{\citenamefont {Discher}\ and\ \citenamefont
  {Eisenberg}(2002)}]{Discher2002}%
  \BibitemOpen
  \bibfield  {author} {\bibinfo {author} {\bibfnamefont {D.~E.}\ \bibnamefont
  {Discher}}\ and\ \bibinfo {author} {\bibfnamefont {A.}~\bibnamefont
  {Eisenberg}},\ }\href {https://www.science.org/doi/10.1126/science.1074972}
  {\bibfield  {journal} {\bibinfo  {journal} {Science}\ }\textbf {\bibinfo
  {volume} {297}},\ \bibinfo {pages} {967} (\bibinfo {year}
  {2002})}\BibitemShut {NoStop}%
\bibitem [{\citenamefont {Jiang}\ \emph {et~al.}(2022)\citenamefont {Jiang},
  \citenamefont {Zhang},\ and\ \citenamefont {Gao}}]{Jiang2022}%
  \BibitemOpen
  \bibfield  {author} {\bibinfo {author} {\bibfnamefont {X.~C.}\ \bibnamefont
  {Jiang}}, \bibinfo {author} {\bibfnamefont {T.}~\bibnamefont {Zhang}},\ and\
  \bibinfo {author} {\bibfnamefont {J.~Q.}\ \bibnamefont {Gao}},\ }\href
  {https://www.sciencedirect.com/science/article/pii/S0169409X22002149}
  {\bibfield  {journal} {\bibinfo  {journal} {Adv. Drug Delivery Rev.}\
  }\textbf {\bibinfo {volume} {187}},\ \bibinfo {pages} {114324} (\bibinfo
  {year} {2022})}\BibitemShut {NoStop}%
\bibitem [{\citenamefont {Wang}\ \emph {et~al.}(2018)\citenamefont {Wang},
  \citenamefont {Dong}, \citenamefont {Li}, \citenamefont {Li}, \citenamefont
  {Cheng}, \citenamefont {Qian}, \citenamefont {Xu}, \citenamefont {Zhang},
  \citenamefont {Hu}, \citenamefont {Chen}, \citenamefont {Du}, \citenamefont
  {Feng}, \citenamefont {Zhao}, \citenamefont {Zhang}, \citenamefont {Liu},
  \citenamefont {Wang}, \citenamefont {Dong}, \citenamefont {Li}, \citenamefont
  {Cheng}, \citenamefont {Qian}, \citenamefont {Xu}, \citenamefont {Zhang},
  \citenamefont {Chen}, \citenamefont {Du}, \citenamefont {Feng}, \citenamefont
  {d~Zhao}, \citenamefont {Zhang}, \citenamefont {f~Liu}, \citenamefont {Li},\
  and\ \citenamefont {Hu}}]{Wang2018}%
  \BibitemOpen
  \bibfield  {author} {\bibinfo {author} {\bibfnamefont {J.}~\bibnamefont
  {Wang}}, \bibinfo {author} {\bibfnamefont {Y.}~\bibnamefont {Dong}}, \bibinfo
  {author} {\bibfnamefont {Y.}~\bibnamefont {Li}}, \bibinfo {author}
  {\bibfnamefont {W.}~\bibnamefont {Li}}, \bibinfo {author} {\bibfnamefont
  {K.}~\bibnamefont {Cheng}}, \bibinfo {author} {\bibfnamefont
  {Y.}~\bibnamefont {Qian}}, \bibinfo {author} {\bibfnamefont {G.}~\bibnamefont
  {Xu}}, \bibinfo {author} {\bibfnamefont {X.}~\bibnamefont {Zhang}}, \bibinfo
  {author} {\bibfnamefont {L.}~\bibnamefont {Hu}}, \bibinfo {author}
  {\bibfnamefont {P.}~\bibnamefont {Chen}}, \bibinfo {author} {\bibfnamefont
  {W.}~\bibnamefont {Du}}, \bibinfo {author} {\bibfnamefont {X.}~\bibnamefont
  {Feng}}, \bibinfo {author} {\bibfnamefont {Y.-D.}\ \bibnamefont {Zhao}},
  \bibinfo {author} {\bibfnamefont {Z.}~\bibnamefont {Zhang}}, \bibinfo
  {author} {\bibfnamefont {B.-F.}\ \bibnamefont {Liu}}, \bibinfo {author}
  {\bibfnamefont {J.}~\bibnamefont {Wang}}, \bibinfo {author} {\bibfnamefont
  {Y.}~\bibnamefont {Dong}}, \bibinfo {author} {\bibfnamefont {W.}~\bibnamefont
  {Li}}, \bibinfo {author} {\bibfnamefont {K.}~\bibnamefont {Cheng}}, \bibinfo
  {author} {\bibfnamefont {Y.}~\bibnamefont {Qian}}, \bibinfo {author}
  {\bibfnamefont {G.}~\bibnamefont {Xu}}, \bibinfo {author} {\bibfnamefont
  {X.}~\bibnamefont {Zhang}}, \bibinfo {author} {\bibfnamefont
  {P.}~\bibnamefont {Chen}}, \bibinfo {author} {\bibfnamefont {W.}~\bibnamefont
  {Du}}, \bibinfo {author} {\bibfnamefont {X.}~\bibnamefont {Feng}}, \bibinfo
  {author} {\bibfnamefont {Y.}~\bibnamefont {d~Zhao}}, \bibinfo {author}
  {\bibfnamefont {Z.}~\bibnamefont {Zhang}}, \bibinfo {author} {\bibfnamefont
  {B.}~\bibnamefont {f~Liu}}, \bibinfo {author} {\bibfnamefont
  {Y.}~\bibnamefont {Li}},\ and\ \bibinfo {author} {\bibfnamefont
  {L.}~\bibnamefont {Hu}},\ }\href
  {https://onlinelibrary.wiley.com/doi/full/10.1002/adfm.201707360} {\bibfield
  {journal} {\bibinfo  {journal} {Adv. Funct. Mater.}\ }\textbf {\bibinfo
  {volume} {28}},\ \bibinfo {pages} {1707360} (\bibinfo {year}
  {2018})}\BibitemShut {NoStop}%
\bibitem [{\citenamefont {Lv}\ \emph {et~al.}(2022)\citenamefont {Lv},
  \citenamefont {Liu}, \citenamefont {Li}, \citenamefont {Lv}, \citenamefont
  {Lu},\ and\ \citenamefont {Xin}}]{Lv2022}%
  \BibitemOpen
  \bibfield  {author} {\bibinfo {author} {\bibfnamefont {W.}~\bibnamefont
  {Lv}}, \bibinfo {author} {\bibfnamefont {Y.}~\bibnamefont {Liu}}, \bibinfo
  {author} {\bibfnamefont {S.}~\bibnamefont {Li}}, \bibinfo {author}
  {\bibfnamefont {L.}~\bibnamefont {Lv}}, \bibinfo {author} {\bibfnamefont
  {H.}~\bibnamefont {Lu}},\ and\ \bibinfo {author} {\bibfnamefont
  {H.}~\bibnamefont {Xin}},\ }\href
  {https://www.ncbi.nlm.nih.gov/pmc/articles/PMC9153106} {\bibfield  {journal}
  {\bibinfo  {journal} {J. Nanobiotechnol.}\ }\textbf {\bibinfo {volume}
  {20}},\ \bibinfo {pages} {248} (\bibinfo {year} {2022})}\BibitemShut
  {NoStop}%
\bibitem [{\citenamefont {Axthelm}\ \emph {et~al.}(2008)\citenamefont
  {Axthelm}, \citenamefont {Casse}, \citenamefont {Koppenol}, \citenamefont
  {Nauser}, \citenamefont {Meier},\ and\ \citenamefont
  {Palivan}}]{Axthelm2008}%
  \BibitemOpen
  \bibfield  {author} {\bibinfo {author} {\bibfnamefont {F.}~\bibnamefont
  {Axthelm}}, \bibinfo {author} {\bibfnamefont {O.}~\bibnamefont {Casse}},
  \bibinfo {author} {\bibfnamefont {W.~H.}\ \bibnamefont {Koppenol}}, \bibinfo
  {author} {\bibfnamefont {T.}~\bibnamefont {Nauser}}, \bibinfo {author}
  {\bibfnamefont {W.}~\bibnamefont {Meier}},\ and\ \bibinfo {author}
  {\bibfnamefont {C.~G.}\ \bibnamefont {Palivan}},\ }\href
  {https://pubs.acs.org/doi/10.1021/jp803032w} {\bibfield  {journal} {\bibinfo
  {journal} {J. Phys. Chem. B}\ }\textbf {\bibinfo {volume} {112}},\ \bibinfo
  {pages} {8211} (\bibinfo {year} {2008})}\BibitemShut {NoStop}%
\bibitem [{\citenamefont {Palivan}\ \emph {et~al.}(2012)\citenamefont
  {Palivan}, \citenamefont {Fischer-Onaca}, \citenamefont {Delcea},
  \citenamefont {Itel},\ and\ \citenamefont {Meier}}]{Palivan2012}%
  \BibitemOpen
  \bibfield  {author} {\bibinfo {author} {\bibfnamefont {C.~G.}\ \bibnamefont
  {Palivan}}, \bibinfo {author} {\bibfnamefont {O.}~\bibnamefont
  {Fischer-Onaca}}, \bibinfo {author} {\bibfnamefont {M.}~\bibnamefont
  {Delcea}}, \bibinfo {author} {\bibfnamefont {F.}~\bibnamefont {Itel}},\ and\
  \bibinfo {author} {\bibfnamefont {W.}~\bibnamefont {Meier}},\ }\href
  {https://pubs.rsc.org/en/content/articlehtml/2012/cs/c1cs15240h} {\bibfield
  {journal} {\bibinfo  {journal} {Chem. Soc. Rev.}\ }\textbf {\bibinfo {volume}
  {41}},\ \bibinfo {pages} {2800} (\bibinfo {year} {2012})}\BibitemShut
  {NoStop}%
\bibitem [{\citenamefont {Gelbart}\ \emph {et~al.}(2012)\citenamefont
  {Gelbart}, \citenamefont {Ben-Shaul},\ and\ \citenamefont
  {Roux}}]{gelbart2012}%
  \BibitemOpen
  \bibfield  {author} {\bibinfo {author} {\bibfnamefont {W.~M.}\ \bibnamefont
  {Gelbart}}, \bibinfo {author} {\bibfnamefont {A.}~\bibnamefont {Ben-Shaul}},\
  and\ \bibinfo {author} {\bibfnamefont {D.}~\bibnamefont {Roux}},\ }\href@noop
  {} {\emph {\bibinfo {title} {Micelles, membranes, microemulsions, and
  monolayers}}}\ (\bibinfo  {publisher} {Springer},\ \bibinfo {year}
  {2012})\BibitemShut {NoStop}%
\bibitem [{\citenamefont {Johnson}\ \emph {et~al.}(2002)\citenamefont
  {Johnson}, \citenamefont {Lewis},\ and\ \citenamefont
  {ALBERTS}}]{johnson2002}%
  \BibitemOpen
  \bibfield  {author} {\bibinfo {author} {\bibfnamefont {A.}~\bibnamefont
  {Johnson}}, \bibinfo {author} {\bibfnamefont {J.}~\bibnamefont {Lewis}},\
  and\ \bibinfo {author} {\bibfnamefont {B.}~\bibnamefont {ALBERTS}},\
  }\href@noop {} {\emph {\bibinfo {title} {Molecular biology of the cell}}}\
  (\bibinfo  {publisher} {Garland Science},\ \bibinfo {year}
  {2002})\BibitemShut {NoStop}%
\bibitem [{\citenamefont {Jahn}\ and\ \citenamefont
  {Südhof}(2003)}]{Jahn2003}%
  \BibitemOpen
  \bibfield  {author} {\bibinfo {author} {\bibfnamefont {R.}~\bibnamefont
  {Jahn}}\ and\ \bibinfo {author} {\bibfnamefont {T.~C.}\ \bibnamefont
  {Südhof}},\ }\href
  {https://www.annualreviews.org/doi/abs/10.1146/annurev.ne.17.030194.001251}
  {\bibfield  {journal} {\bibinfo  {journal} {Annu. Rev. Neurosci.}\ }\textbf
  {\bibinfo {volume} {17}},\ \bibinfo {pages} {219} (\bibinfo {year}
  {2003})}\BibitemShut {NoStop}%
\bibitem [{\citenamefont {Jahn}\ and\ \citenamefont
  {Fasshauer}(2012)}]{Jahn2012}%
  \BibitemOpen
  \bibfield  {author} {\bibinfo {author} {\bibfnamefont {R.}~\bibnamefont
  {Jahn}}\ and\ \bibinfo {author} {\bibfnamefont {D.}~\bibnamefont
  {Fasshauer}},\ }\href {https://www.nature.com/articles/nature11320}
  {\bibfield  {journal} {\bibinfo  {journal} {Nature}\ }\textbf {\bibinfo
  {volume} {490}},\ \bibinfo {pages} {201} (\bibinfo {year}
  {2012})}\BibitemShut {NoStop}%
\bibitem [{\citenamefont {Düzgüneş}\ and\ \citenamefont
  {Nir}(1999)}]{Düzgüneş1999}%
  \BibitemOpen
  \bibfield  {author} {\bibinfo {author} {\bibfnamefont {N.}~\bibnamefont
  {Düzgüneş}}\ and\ \bibinfo {author} {\bibfnamefont {S.}~\bibnamefont
  {Nir}},\ }\href
  {https://www.sciencedirect.com/science/article/pii/S0169409X9900037X}
  {\bibfield  {journal} {\bibinfo  {journal} {Adv. Drug Delivery Rev.}\
  }\textbf {\bibinfo {volume} {40}},\ \bibinfo {pages} {3} (\bibinfo {year}
  {1999})}\BibitemShut {NoStop}%
\bibitem [{\citenamefont {Poon}\ \emph {et~al.}(2020)\citenamefont {Poon},
  \citenamefont {Kingston}, \citenamefont {Ouyang}, \citenamefont {Ngo},\ and\
  \citenamefont {Chan}}]{Poon2020}%
  \BibitemOpen
  \bibfield  {author} {\bibinfo {author} {\bibfnamefont {W.}~\bibnamefont
  {Poon}}, \bibinfo {author} {\bibfnamefont {B.~R.}\ \bibnamefont {Kingston}},
  \bibinfo {author} {\bibfnamefont {B.}~\bibnamefont {Ouyang}}, \bibinfo
  {author} {\bibfnamefont {W.}~\bibnamefont {Ngo}},\ and\ \bibinfo {author}
  {\bibfnamefont {W.~C.}\ \bibnamefont {Chan}},\ }\href
  {https://www.nature.com/articles/s41565-020-0759-5} {\bibfield  {journal}
  {\bibinfo  {journal} {Nat. Nanotechnol.}\ }\textbf {\bibinfo {volume} {15}},\
  \bibinfo {pages} {819} (\bibinfo {year} {2020})}\BibitemShut {NoStop}%
\bibitem [{\citenamefont {Shen}\ \emph {et~al.}(2019)\citenamefont {Shen},
  \citenamefont {Ye}, \citenamefont {Kröger}, \citenamefont {Tang},\ and\
  \citenamefont {Li}}]{Shen2019}%
  \BibitemOpen
  \bibfield  {author} {\bibinfo {author} {\bibfnamefont {Z.}~\bibnamefont
  {Shen}}, \bibinfo {author} {\bibfnamefont {H.}~\bibnamefont {Ye}}, \bibinfo
  {author} {\bibfnamefont {M.}~\bibnamefont {Kröger}}, \bibinfo {author}
  {\bibfnamefont {S.}~\bibnamefont {Tang}},\ and\ \bibinfo {author}
  {\bibfnamefont {Y.}~\bibnamefont {Li}},\ }\href
  {http://dx.doi.org/10.1039/C9NR02408E} {\bibfield  {journal} {\bibinfo
  {journal} {Nanoscale}\ }\textbf {\bibinfo {volume} {11}},\ \bibinfo {pages}
  {15971} (\bibinfo {year} {2019})}\BibitemShut {NoStop}%
\bibitem [{\citenamefont {Ho}\ \emph {et~al.}(2021)\citenamefont {Ho},
  \citenamefont {Siggel}, \citenamefont {Camacho}, \citenamefont {Bhaskara},
  \citenamefont {Hicks}, \citenamefont {Yao}, \citenamefont {Zhang},
  \citenamefont {Köfinger}, \citenamefont {Hummer},\ and\ \citenamefont
  {Noy}}]{Ho2021}%
  \BibitemOpen
  \bibfield  {author} {\bibinfo {author} {\bibfnamefont {N.~T.}\ \bibnamefont
  {Ho}}, \bibinfo {author} {\bibfnamefont {M.}~\bibnamefont {Siggel}}, \bibinfo
  {author} {\bibfnamefont {K.~V.}\ \bibnamefont {Camacho}}, \bibinfo {author}
  {\bibfnamefont {R.~M.}\ \bibnamefont {Bhaskara}}, \bibinfo {author}
  {\bibfnamefont {J.~M.}\ \bibnamefont {Hicks}}, \bibinfo {author}
  {\bibfnamefont {Y.-C.}\ \bibnamefont {Yao}}, \bibinfo {author} {\bibfnamefont
  {Y.}~\bibnamefont {Zhang}}, \bibinfo {author} {\bibfnamefont
  {J.}~\bibnamefont {Köfinger}}, \bibinfo {author} {\bibfnamefont
  {G.}~\bibnamefont {Hummer}},\ and\ \bibinfo {author} {\bibfnamefont
  {A.}~\bibnamefont {Noy}},\ }\href
  {https://www.pnas.org/doi/abs/10.1073/pnas.2016974118} {\bibfield  {journal}
  {\bibinfo  {journal} {Proc. Natl. Acad. Sci. U.S.A.}\ }\textbf {\bibinfo
  {volume} {118}},\ \bibinfo {pages} {e2016974118} (\bibinfo {year}
  {2021})}\BibitemShut {NoStop}%
\bibitem [{\citenamefont {Luo}\ and\ \citenamefont
  {Eisenberg}(2001)}]{Luo2001}%
  \BibitemOpen
  \bibfield  {author} {\bibinfo {author} {\bibfnamefont {L.}~\bibnamefont
  {Luo}}\ and\ \bibinfo {author} {\bibfnamefont {A.}~\bibnamefont
  {Eisenberg}},\ }\href {https://pubs.acs.org/doi/full/10.1021/la0104370}
  {\bibfield  {journal} {\bibinfo  {journal} {Langmuir}\ }\textbf {\bibinfo
  {volume} {17}},\ \bibinfo {pages} {6804} (\bibinfo {year}
  {2001})}\BibitemShut {NoStop}%
\bibitem [{\citenamefont {Smeijers}\ \emph {et~al.}(2006)\citenamefont
  {Smeijers}, \citenamefont {Markvoort}, \citenamefont {Pieterse},\ and\
  \citenamefont {Hilbers}}]{Smeijers2006}%
  \BibitemOpen
  \bibfield  {author} {\bibinfo {author} {\bibfnamefont {A.~F.}\ \bibnamefont
  {Smeijers}}, \bibinfo {author} {\bibfnamefont {A.~J.}\ \bibnamefont
  {Markvoort}}, \bibinfo {author} {\bibfnamefont {K.}~\bibnamefont
  {Pieterse}},\ and\ \bibinfo {author} {\bibfnamefont {P.~A.~J.}\ \bibnamefont
  {Hilbers}},\ }\href {https://pubs.acs.org/doi/full/10.1021/jp060824o}
  {\bibfield  {journal} {\bibinfo  {journal} {J. Phys. Chem. B}\ }\textbf
  {\bibinfo {volume} {110}},\ \bibinfo {pages} {13212} (\bibinfo {year}
  {2006})}\BibitemShut {NoStop}%
\bibitem [{\citenamefont {Markvoort}\ \emph {et~al.}(2007)\citenamefont
  {Markvoort}, \citenamefont {Smeijers}, \citenamefont {Pieterse},
  \citenamefont {Santen},\ and\ \citenamefont {Hilbers}}]{Markvoort2007}%
  \BibitemOpen
  \bibfield  {author} {\bibinfo {author} {\bibfnamefont {A.~J.}\ \bibnamefont
  {Markvoort}}, \bibinfo {author} {\bibfnamefont {A.~F.}\ \bibnamefont
  {Smeijers}}, \bibinfo {author} {\bibfnamefont {K.}~\bibnamefont {Pieterse}},
  \bibinfo {author} {\bibfnamefont {R.~A.~V.}\ \bibnamefont {Santen}},\ and\
  \bibinfo {author} {\bibfnamefont {P.~A.}\ \bibnamefont {Hilbers}},\ }\href
  {https://pubs.acs.org/doi/full/10.1021/jp068277u} {\bibfield  {journal}
  {\bibinfo  {journal} {J. Phys. Chem. B}\ }\textbf {\bibinfo {volume} {111}},\
  \bibinfo {pages} {5719} (\bibinfo {year} {2007})}\BibitemShut {NoStop}%
\bibitem [{\citenamefont {Markin}\ \emph {et~al.}(1984)\citenamefont {Markin},
  \citenamefont {Kozlov},\ and\ \citenamefont {Borovjagin}}]{Markin1984}%
  \BibitemOpen
  \bibfield  {author} {\bibinfo {author} {\bibfnamefont {V.~S.}\ \bibnamefont
  {Markin}}, \bibinfo {author} {\bibfnamefont {M.~M.}\ \bibnamefont {Kozlov}},\
  and\ \bibinfo {author} {\bibfnamefont {V.~L.}\ \bibnamefont {Borovjagin}},\
  }\href {https://pubmed.ncbi.nlm.nih.gov/6510702} {\bibfield  {journal}
  {\bibinfo  {journal} {Gen. Physiol. Biophys.}\ }\textbf {\bibinfo {volume}
  {3}},\ \bibinfo {pages} {361} (\bibinfo {year} {1984})}\BibitemShut {NoStop}%
\bibitem [{\citenamefont {Chernomordik}\ \emph {et~al.}(1987)\citenamefont
  {Chernomordik}, \citenamefont {Melikyan},\ and\ \citenamefont
  {Chizmadzhev}}]{Chernomordik1987}%
  \BibitemOpen
  \bibfield  {author} {\bibinfo {author} {\bibfnamefont {L.~V.}\ \bibnamefont
  {Chernomordik}}, \bibinfo {author} {\bibfnamefont {G.~B.}\ \bibnamefont
  {Melikyan}},\ and\ \bibinfo {author} {\bibfnamefont {Y.~A.}\ \bibnamefont
  {Chizmadzhev}},\ }\href {https://pubmed.ncbi.nlm.nih.gov/3307918/} {\bibfield
   {journal} {\bibinfo  {journal} {Biochim. Biophys. Acta, Rev. Biomembr.}\
  }\textbf {\bibinfo {volume} {906}},\ \bibinfo {pages} {309} (\bibinfo {year}
  {1987})}\BibitemShut {NoStop}%
\bibitem [{\citenamefont {Chernomordik}\ \emph {et~al.}(1995)\citenamefont
  {Chernomordik}, \citenamefont {Kozlov},\ and\ \citenamefont
  {Zimmerberg}}]{Chernomordik1995}%
  \BibitemOpen
  \bibfield  {author} {\bibinfo {author} {\bibfnamefont {L.}~\bibnamefont
  {Chernomordik}}, \bibinfo {author} {\bibfnamefont {M.~M.}\ \bibnamefont
  {Kozlov}},\ and\ \bibinfo {author} {\bibfnamefont {J.}~\bibnamefont
  {Zimmerberg}},\ }\href {https://link.springer.com/article/10.1007/BF00232676}
  {\bibfield  {journal} {\bibinfo  {journal} {J. Membr. Biol.}\ }\textbf
  {\bibinfo {volume} {146}},\ \bibinfo {pages} {1} (\bibinfo {year}
  {1995})}\BibitemShut {NoStop}%
\bibitem [{\citenamefont {Grafmüller}\ \emph {et~al.}(2007)\citenamefont
  {Grafmüller}, \citenamefont {Shillcock},\ and\ \citenamefont
  {Lipowsky}}]{Grafmüller2007}%
  \BibitemOpen
  \bibfield  {author} {\bibinfo {author} {\bibfnamefont {A.}~\bibnamefont
  {Grafmüller}}, \bibinfo {author} {\bibfnamefont {J.}~\bibnamefont
  {Shillcock}},\ and\ \bibinfo {author} {\bibfnamefont {R.}~\bibnamefont
  {Lipowsky}},\ }\href
  {https://journals.aps.org/prl/abstract/10.1103/PhysRevLett.98.218101}
  {\bibfield  {journal} {\bibinfo  {journal} {Phys. Rev. Lett.}\ }\textbf
  {\bibinfo {volume} {98}},\ \bibinfo {pages} {218101} (\bibinfo {year}
  {2007})}\BibitemShut {NoStop}%
\bibitem [{\citenamefont {Gao}\ \emph {et~al.}(2008)\citenamefont {Gao},
  \citenamefont {Lipowsky},\ and\ \citenamefont {Shillcock}}]{Gao2008}%
  \BibitemOpen
  \bibfield  {author} {\bibinfo {author} {\bibfnamefont {L.}~\bibnamefont
  {Gao}}, \bibinfo {author} {\bibfnamefont {R.}~\bibnamefont {Lipowsky}},\ and\
  \bibinfo {author} {\bibfnamefont {J.}~\bibnamefont {Shillcock}},\ }\href
  {https://pubs.rsc.org/en/content/articlehtml/2008/sm/b801407h} {\bibfield
  {journal} {\bibinfo  {journal} {Soft Matter}\ }\textbf {\bibinfo {volume}
  {4}},\ \bibinfo {pages} {1208} (\bibinfo {year} {2008})}\BibitemShut
  {NoStop}%
\bibitem [{\citenamefont {Li}\ \emph {et~al.}(2009)\citenamefont {Li},
  \citenamefont {Liu}, \citenamefont {Wang}, \citenamefont {Deng},\ and\
  \citenamefont {Liang}}]{Li2009}%
  \BibitemOpen
  \bibfield  {author} {\bibinfo {author} {\bibfnamefont {X.}~\bibnamefont
  {Li}}, \bibinfo {author} {\bibfnamefont {Y.}~\bibnamefont {Liu}}, \bibinfo
  {author} {\bibfnamefont {L.}~\bibnamefont {Wang}}, \bibinfo {author}
  {\bibfnamefont {M.}~\bibnamefont {Deng}},\ and\ \bibinfo {author}
  {\bibfnamefont {H.}~\bibnamefont {Liang}},\ }\href
  {https://pubs.rsc.org/en/content/articlehtml/2009/cp/b817773bb} {\bibfield
  {journal} {\bibinfo  {journal} {Phys. Chem. Chem. Phys.}\ }\textbf {\bibinfo
  {volume} {11}},\ \bibinfo {pages} {4051} (\bibinfo {year}
  {2009})}\BibitemShut {NoStop}%
\bibitem [{\citenamefont {Risselada}\ \emph {et~al.}(2012)\citenamefont
  {Risselada}, \citenamefont {Marelli}, \citenamefont {Fuhrmans}, \citenamefont
  {Smirnova}, \citenamefont {Grubmüller}, \citenamefont {Marrink},\ and\
  \citenamefont {Müller}}]{Risselada2012}%
  \BibitemOpen
  \bibfield  {author} {\bibinfo {author} {\bibfnamefont {H.~J.}\ \bibnamefont
  {Risselada}}, \bibinfo {author} {\bibfnamefont {G.}~\bibnamefont {Marelli}},
  \bibinfo {author} {\bibfnamefont {M.}~\bibnamefont {Fuhrmans}}, \bibinfo
  {author} {\bibfnamefont {Y.~G.}\ \bibnamefont {Smirnova}}, \bibinfo {author}
  {\bibfnamefont {H.}~\bibnamefont {Grubmüller}}, \bibinfo {author}
  {\bibfnamefont {S.~J.}\ \bibnamefont {Marrink}},\ and\ \bibinfo {author}
  {\bibfnamefont {M.}~\bibnamefont {Müller}},\ }\href
  {https://journals.plos.org/plosone/article?id=10.1371/journal.pone.0038302}
  {\bibfield  {journal} {\bibinfo  {journal} {PLoS One}\ }\textbf {\bibinfo
  {volume} {7}},\ \bibinfo {pages} {e38302} (\bibinfo {year}
  {2012})}\BibitemShut {NoStop}%
\bibitem [{\citenamefont {Marrink}\ and\ \citenamefont
  {Mark}(2003)}]{Marrink2003}%
  \BibitemOpen
  \bibfield  {author} {\bibinfo {author} {\bibfnamefont {S.~J.}\ \bibnamefont
  {Marrink}}\ and\ \bibinfo {author} {\bibfnamefont {A.~E.}\ \bibnamefont
  {Mark}},\ }\href {https://doi.org/10.1021/ja036138+} {\bibfield  {journal}
  {\bibinfo  {journal} {J. Am. Chem. Soc.}\ }\textbf {\bibinfo {volume}
  {125}},\ \bibinfo {pages} {11144} (\bibinfo {year} {2003})}\BibitemShut
  {NoStop}%
\bibitem [{\citenamefont {Shillcock}\ and\ \citenamefont
  {Lipowsky}(2005)}]{Shillcock2005}%
  \BibitemOpen
  \bibfield  {author} {\bibinfo {author} {\bibfnamefont {J.~C.}\ \bibnamefont
  {Shillcock}}\ and\ \bibinfo {author} {\bibfnamefont {R.}~\bibnamefont
  {Lipowsky}},\ }\href {https://www.nature.com/articles/nmat1333} {\bibfield
  {journal} {\bibinfo  {journal} {Nat. Mater.}\ }\textbf {\bibinfo {volume}
  {4}},\ \bibinfo {pages} {225} (\bibinfo {year} {2005})}\BibitemShut {NoStop}%
\bibitem [{\citenamefont {Siegel}(1999)}]{Siegel1999}%
  \BibitemOpen
  \bibfield  {author} {\bibinfo {author} {\bibfnamefont {D.~P.}\ \bibnamefont
  {Siegel}},\ }\href
  {https://www.sciencedirect.com/science/article/pii/S0006349599771973}
  {\bibfield  {journal} {\bibinfo  {journal} {Biophys. J.}\ }\textbf {\bibinfo
  {volume} {76}},\ \bibinfo {pages} {291} (\bibinfo {year} {1999})}\BibitemShut
  {NoStop}%
\bibitem [{\citenamefont {Siegel}(1993)}]{Siegel1993}%
  \BibitemOpen
  \bibfield  {author} {\bibinfo {author} {\bibfnamefont {D.~P.}\ \bibnamefont
  {Siegel}},\ }\href
  {https://www.sciencedirect.com/science/article/pii/S0006349593812566}
  {\bibfield  {journal} {\bibinfo  {journal} {Biophys. J.}\ }\textbf {\bibinfo
  {volume} {65}},\ \bibinfo {pages} {2124} (\bibinfo {year}
  {1993})}\BibitemShut {NoStop}%
\bibitem [{\citenamefont {Noguchi}\ and\ \citenamefont
  {Takasu}(2001)}]{Noguchi2001}%
  \BibitemOpen
  \bibfield  {author} {\bibinfo {author} {\bibfnamefont {H.}~\bibnamefont
  {Noguchi}}\ and\ \bibinfo {author} {\bibfnamefont {M.}~\bibnamefont
  {Takasu}},\ }\href {https://aip.scitation.org/doi/abs/10.1063/1.1414314}
  {\bibfield  {journal} {\bibinfo  {journal} {J. Chem. Phys.}\ }\textbf
  {\bibinfo {volume} {115}},\ \bibinfo {pages} {9547} (\bibinfo {year}
  {2001})}\BibitemShut {NoStop}%
\bibitem [{\citenamefont {Matsuoka}\ \emph {et~al.}(1998)\citenamefont
  {Matsuoka}, \citenamefont {Orci}, \citenamefont {Amherdt}, \citenamefont
  {Bednarek}, \citenamefont {Hamamoto}, \citenamefont {Schekman},\ and\
  \citenamefont {Yeung}}]{Matsuoka1998}%
  \BibitemOpen
  \bibfield  {author} {\bibinfo {author} {\bibfnamefont {K.}~\bibnamefont
  {Matsuoka}}, \bibinfo {author} {\bibfnamefont {L.}~\bibnamefont {Orci}},
  \bibinfo {author} {\bibfnamefont {M.}~\bibnamefont {Amherdt}}, \bibinfo
  {author} {\bibfnamefont {S.~Y.}\ \bibnamefont {Bednarek}}, \bibinfo {author}
  {\bibfnamefont {S.}~\bibnamefont {Hamamoto}}, \bibinfo {author}
  {\bibfnamefont {R.}~\bibnamefont {Schekman}},\ and\ \bibinfo {author}
  {\bibfnamefont {T.}~\bibnamefont {Yeung}},\ }\href
  {https://pubmed.ncbi.nlm.nih.gov/9568718/} {\bibfield  {journal} {\bibinfo
  {journal} {Cell}\ }\textbf {\bibinfo {volume} {93}},\ \bibinfo {pages} {263}
  (\bibinfo {year} {1998})}\BibitemShut {NoStop}%
\bibitem [{\citenamefont {Spang}\ \emph {et~al.}(1998)\citenamefont {Spang},
  \citenamefont {Matsuoka}, \citenamefont {Hamamoto}, \citenamefont
  {Schekman},\ and\ \citenamefont {Orci}}]{Spang1998}%
  \BibitemOpen
  \bibfield  {author} {\bibinfo {author} {\bibfnamefont {A.}~\bibnamefont
  {Spang}}, \bibinfo {author} {\bibfnamefont {K.}~\bibnamefont {Matsuoka}},
  \bibinfo {author} {\bibfnamefont {S.}~\bibnamefont {Hamamoto}}, \bibinfo
  {author} {\bibfnamefont {R.}~\bibnamefont {Schekman}},\ and\ \bibinfo
  {author} {\bibfnamefont {L.}~\bibnamefont {Orci}},\ }\href
  {https://www.pnas.org/doi/abs/10.1073/pnas.95.19.11199} {\bibfield  {journal}
  {\bibinfo  {journal} {Proc. Natl. Acad. Sci. U.S.A.}\ }\textbf {\bibinfo
  {volume} {95}},\ \bibinfo {pages} {11199} (\bibinfo {year}
  {1998})}\BibitemShut {NoStop}%
\bibitem [{\citenamefont {Takei}\ \emph {et~al.}(1998)\citenamefont {Takei},
  \citenamefont {Haucke}, \citenamefont {Slepnev}, \citenamefont {Farsad},
  \citenamefont {Salazar}, \citenamefont {Chen},\ and\ \citenamefont
  {Camilli}}]{Takei1998}%
  \BibitemOpen
  \bibfield  {author} {\bibinfo {author} {\bibfnamefont {K.}~\bibnamefont
  {Takei}}, \bibinfo {author} {\bibfnamefont {V.}~\bibnamefont {Haucke}},
  \bibinfo {author} {\bibfnamefont {V.}~\bibnamefont {Slepnev}}, \bibinfo
  {author} {\bibfnamefont {K.}~\bibnamefont {Farsad}}, \bibinfo {author}
  {\bibfnamefont {M.}~\bibnamefont {Salazar}}, \bibinfo {author} {\bibfnamefont
  {H.}~\bibnamefont {Chen}},\ and\ \bibinfo {author} {\bibfnamefont {P.~D.}\
  \bibnamefont {Camilli}},\ }\href
  {http://www.cell.com/article/S0092867400812283} {\bibfield  {journal}
  {\bibinfo  {journal} {Cell}\ }\textbf {\bibinfo {volume} {94}},\ \bibinfo
  {pages} {131} (\bibinfo {year} {1998})}\BibitemShut {NoStop}%
\bibitem [{\citenamefont {Kozlovsky}\ and\ \citenamefont
  {Kozlov}(2003)}]{Kozlovsky2003}%
  \BibitemOpen
  \bibfield  {author} {\bibinfo {author} {\bibfnamefont {Y.}~\bibnamefont
  {Kozlovsky}}\ and\ \bibinfo {author} {\bibfnamefont {M.~M.}\ \bibnamefont
  {Kozlov}},\ }\href
  {https://www.sciencedirect.com/science/article/pii/S0006349503744579}
  {\bibfield  {journal} {\bibinfo  {journal} {Biophys. J.}\ }\textbf {\bibinfo
  {volume} {85}},\ \bibinfo {pages} {85} (\bibinfo {year} {2003})}\BibitemShut
  {NoStop}%
\bibitem [{\citenamefont {Shemesh}\ \emph {et~al.}(2003)\citenamefont
  {Shemesh}, \citenamefont {Luini}, \citenamefont {Malhotra}, \citenamefont
  {Burger},\ and\ \citenamefont {Kozlov}}]{Shemesh2003}%
  \BibitemOpen
  \bibfield  {author} {\bibinfo {author} {\bibfnamefont {T.}~\bibnamefont
  {Shemesh}}, \bibinfo {author} {\bibfnamefont {A.}~\bibnamefont {Luini}},
  \bibinfo {author} {\bibfnamefont {V.}~\bibnamefont {Malhotra}}, \bibinfo
  {author} {\bibfnamefont {K.~N.}\ \bibnamefont {Burger}},\ and\ \bibinfo
  {author} {\bibfnamefont {M.~M.}\ \bibnamefont {Kozlov}},\ }\href
  {https://www.sciencedirect.com/science/article/pii/S0006349503747961}
  {\bibfield  {journal} {\bibinfo  {journal} {Biophys. J.}\ }\textbf {\bibinfo
  {volume} {85}},\ \bibinfo {pages} {3813} (\bibinfo {year}
  {2003})}\BibitemShut {NoStop}%
\bibitem [{\citenamefont {Zimmerberg}\ and\ \citenamefont
  {Kozlov}(2005)}]{Zimmerberg2005}%
  \BibitemOpen
  \bibfield  {author} {\bibinfo {author} {\bibfnamefont {J.}~\bibnamefont
  {Zimmerberg}}\ and\ \bibinfo {author} {\bibfnamefont {M.~M.}\ \bibnamefont
  {Kozlov}},\ }\href {https://www.nature.com/articles/nrm1784} {\bibfield
  {journal} {\bibinfo  {journal} {Nat. Rev. Mol. Cell Biol.}\ }\textbf
  {\bibinfo {volume} {7}},\ \bibinfo {pages} {9} (\bibinfo {year}
  {2005})}\BibitemShut {NoStop}%
\bibitem [{\citenamefont {Atilgan}\ and\ \citenamefont
  {Sun}(2007)}]{Atilgan2007}%
  \BibitemOpen
  \bibfield  {author} {\bibinfo {author} {\bibfnamefont {E.}~\bibnamefont
  {Atilgan}}\ and\ \bibinfo {author} {\bibfnamefont {S.~X.}\ \bibnamefont
  {Sun}},\ }\href {https://pubmed.ncbi.nlm.nih.gov/17362130/} {\bibfield
  {journal} {\bibinfo  {journal} {J. Chem. Phys.}\ }\textbf {\bibinfo {volume}
  {126}},\ \bibinfo {pages} {095102} (\bibinfo {year} {2007})}\BibitemShut
  {NoStop}%
\bibitem [{\citenamefont {Campelo}\ \emph {et~al.}(2008)\citenamefont
  {Campelo}, \citenamefont {Mc.m.ahon},\ and\ \citenamefont
  {Kozlov}}]{Campelo2008}%
  \BibitemOpen
  \bibfield  {author} {\bibinfo {author} {\bibfnamefont {F.}~\bibnamefont
  {Campelo}}, \bibinfo {author} {\bibfnamefont {H.~T.}\ \bibnamefont
  {Mc.m.ahon}},\ and\ \bibinfo {author} {\bibfnamefont {M.~M.}\ \bibnamefont
  {Kozlov}},\ }\href {https://pubmed.ncbi.nlm.nih.gov/18515373/} {\bibfield
  {journal} {\bibinfo  {journal} {Biophys. J.}\ }\textbf {\bibinfo {volume}
  {95}},\ \bibinfo {pages} {2325} (\bibinfo {year} {2008})}\BibitemShut
  {NoStop}%
\bibitem [{\citenamefont {Kozlov}\ \emph {et~al.}(2010)\citenamefont {Kozlov},
  \citenamefont {Mc.m.ahon},\ and\ \citenamefont {Chernomordik}}]{Kozlov2010}%
  \BibitemOpen
  \bibfield  {author} {\bibinfo {author} {\bibfnamefont {M.~M.}\ \bibnamefont
  {Kozlov}}, \bibinfo {author} {\bibfnamefont {H.~T.}\ \bibnamefont
  {Mc.m.ahon}},\ and\ \bibinfo {author} {\bibfnamefont {L.~V.}\ \bibnamefont
  {Chernomordik}},\ }\href
  {https://www.sciencedirect.com/science/article/pii/S0968000410001155}
  {\bibfield  {journal} {\bibinfo  {journal} {Trends Biochem. Sci.}\ }\textbf
  {\bibinfo {volume} {35}},\ \bibinfo {pages} {699} (\bibinfo {year}
  {2010})}\BibitemShut {NoStop}%
\bibitem [{\citenamefont {Kozlov}\ and\ \citenamefont
  {Taraska}(2022)}]{Kozlov2022}%
  \BibitemOpen
  \bibfield  {author} {\bibinfo {author} {\bibfnamefont {M.~M.}\ \bibnamefont
  {Kozlov}}\ and\ \bibinfo {author} {\bibfnamefont {J.~W.}\ \bibnamefont
  {Taraska}},\ }\href {https://www.nature.com/articles/s41580-022-00511-9}
  {\bibfield  {journal} {\bibinfo  {journal} {Nat. Rev. Mol. Cell Biol.}\ ,\
  \bibinfo {pages} {1}} (\bibinfo {year} {2022})}\BibitemShut {NoStop}%
\bibitem [{\citenamefont {Yamamoto}\ and\ \citenamefont
  {Hyodo}(2003)}]{Yamamoto2003}%
  \BibitemOpen
  \bibfield  {author} {\bibinfo {author} {\bibfnamefont {S.}~\bibnamefont
  {Yamamoto}}\ and\ \bibinfo {author} {\bibfnamefont {S.~A.}\ \bibnamefont
  {Hyodo}},\ }\href {https://aip.scitation.org/doi/abs/10.1063/1.1563613}
  {\bibfield  {journal} {\bibinfo  {journal} {J. Chem. Phys.}\ }\textbf
  {\bibinfo {volume} {118}},\ \bibinfo {pages} {7937} (\bibinfo {year}
  {2003})}\BibitemShut {NoStop}%
\bibitem [{\citenamefont {Park}\ \emph {et~al.}(2019)\citenamefont {Park},
  \citenamefont {Yang}, \citenamefont {Li}, \citenamefont {Deng}, \citenamefont
  {Zhu}, \citenamefont {Young}, \citenamefont {Ericsson}, \citenamefont
  {Andringa}, \citenamefont {Minnaard}, \citenamefont {Zhu}, \citenamefont
  {Sun}, \citenamefont {Moody}, \citenamefont {Morris}, \citenamefont {Fan},\
  and\ \citenamefont {Hsu}}]{Park2019}%
  \BibitemOpen
  \bibfield  {author} {\bibinfo {author} {\bibfnamefont {S.~Y.}\ \bibnamefont
  {Park}}, \bibinfo {author} {\bibfnamefont {J.~S.}\ \bibnamefont {Yang}},
  \bibinfo {author} {\bibfnamefont {Z.}~\bibnamefont {Li}}, \bibinfo {author}
  {\bibfnamefont {P.}~\bibnamefont {Deng}}, \bibinfo {author} {\bibfnamefont
  {X.}~\bibnamefont {Zhu}}, \bibinfo {author} {\bibfnamefont {D.}~\bibnamefont
  {Young}}, \bibinfo {author} {\bibfnamefont {M.}~\bibnamefont {Ericsson}},
  \bibinfo {author} {\bibfnamefont {R.~L.}\ \bibnamefont {Andringa}}, \bibinfo
  {author} {\bibfnamefont {A.~J.}\ \bibnamefont {Minnaard}}, \bibinfo {author}
  {\bibfnamefont {C.}~\bibnamefont {Zhu}}, \bibinfo {author} {\bibfnamefont
  {F.}~\bibnamefont {Sun}}, \bibinfo {author} {\bibfnamefont {D.~B.}\
  \bibnamefont {Moody}}, \bibinfo {author} {\bibfnamefont {A.~J.}\ \bibnamefont
  {Morris}}, \bibinfo {author} {\bibfnamefont {J.}~\bibnamefont {Fan}},\ and\
  \bibinfo {author} {\bibfnamefont {V.~W.}\ \bibnamefont {Hsu}},\ }\href
  {https://www.nature.com/articles/s41467-019-11324-4} {\bibfield  {journal}
  {\bibinfo  {journal} {Nat. Commun.}\ }\textbf {\bibinfo {volume} {10}},\
  \bibinfo {pages} {1} (\bibinfo {year} {2019})}\BibitemShut {NoStop}%
\bibitem [{\citenamefont {Katsov}\ \emph {et~al.}(2004)\citenamefont {Katsov},
  \citenamefont {Müller},\ and\ \citenamefont {Schick}}]{Katsov2004}%
  \BibitemOpen
  \bibfield  {author} {\bibinfo {author} {\bibfnamefont {K.}~\bibnamefont
  {Katsov}}, \bibinfo {author} {\bibfnamefont {M.}~\bibnamefont {Müller}},\
  and\ \bibinfo {author} {\bibfnamefont {M.}~\bibnamefont {Schick}},\ }\href
  {https://doi.org/10.1529/biophysj.103.038943} {\bibfield  {journal} {\bibinfo
   {journal} {Biophys. J.}\ }\textbf {\bibinfo {volume} {87}},\ \bibinfo
  {pages} {3277} (\bibinfo {year} {2004})}\BibitemShut {NoStop}%
\bibitem [{\citenamefont {Katsov}\ \emph {et~al.}(2006)\citenamefont {Katsov},
  \citenamefont {Müller},\ and\ \citenamefont {Schickz}}]{Katsov2006}%
  \BibitemOpen
  \bibfield  {author} {\bibinfo {author} {\bibfnamefont {K.}~\bibnamefont
  {Katsov}}, \bibinfo {author} {\bibfnamefont {M.}~\bibnamefont {Müller}},\
  and\ \bibinfo {author} {\bibfnamefont {M.}~\bibnamefont {Schickz}},\ }\href
  {http://www.cell.com/article/S0006349506722792} {\bibfield  {journal}
  {\bibinfo  {journal} {Biophys. J.}\ }\textbf {\bibinfo {volume} {90}},\
  \bibinfo {pages} {915} (\bibinfo {year} {2006})}\BibitemShut {NoStop}%
\bibitem [{\citenamefont {Lee}\ and\ \citenamefont {Schick}(2007)}]{Lee2007}%
  \BibitemOpen
  \bibfield  {author} {\bibinfo {author} {\bibfnamefont {J.~Y.}\ \bibnamefont
  {Lee}}\ and\ \bibinfo {author} {\bibfnamefont {M.}~\bibnamefont {Schick}},\
  }\href {https://www.sciencedirect.com/science/article/pii/S0006349507711931}
  {\bibfield  {journal} {\bibinfo  {journal} {Biophys. J.}\ }\textbf {\bibinfo
  {volume} {92}},\ \bibinfo {pages} {3938} (\bibinfo {year}
  {2007})}\BibitemShut {NoStop}%
\bibitem [{\citenamefont {Lee}\ and\ \citenamefont {Schick}(2008)}]{Lee2008}%
  \BibitemOpen
  \bibfield  {author} {\bibinfo {author} {\bibfnamefont {J.}~\bibnamefont
  {Lee}}\ and\ \bibinfo {author} {\bibfnamefont {M.}~\bibnamefont {Schick}},\
  }\href {https://www.sciencedirect.com/science/article/pii/S0006349508706088}
  {\bibfield  {journal} {\bibinfo  {journal} {Biophys. J.}\ }\textbf {\bibinfo
  {volume} {94}},\ \bibinfo {pages} {1699} (\bibinfo {year}
  {2008})}\BibitemShut {NoStop}%
\bibitem [{\citenamefont {Duan}\ \emph {et~al.}(2022)\citenamefont {Duan},
  \citenamefont {Li},\ and\ \citenamefont {Wang}}]{Duan2022}%
  \BibitemOpen
  \bibfield  {author} {\bibinfo {author} {\bibfnamefont {C.}~\bibnamefont
  {Duan}}, \bibinfo {author} {\bibfnamefont {W.}~\bibnamefont {Li}},\ and\
  \bibinfo {author} {\bibfnamefont {R.}~\bibnamefont {Wang}},\ }\href
  {https://pubs.acs.org/doi/full/10.1021/acs.macromol.1c02289} {\bibfield
  {journal} {\bibinfo  {journal} {Macromolecules}\ }\textbf {\bibinfo {volume}
  {55}},\ \bibinfo {pages} {906} (\bibinfo {year} {2022})}\BibitemShut
  {NoStop}%
\bibitem [{\citenamefont {Nakamura}\ and\ \citenamefont
  {Wang}(2012)}]{Issei2012}%
  \BibitemOpen
  \bibfield  {author} {\bibinfo {author} {\bibfnamefont {I.}~\bibnamefont
  {Nakamura}}\ and\ \bibinfo {author} {\bibfnamefont {Z.-G.}\ \bibnamefont
  {Wang}},\ }\href {http://dx.doi.org/10.1039/C2SM25606A} {\bibfield  {journal}
  {\bibinfo  {journal} {Soft Matter}\ }\textbf {\bibinfo {volume} {8}},\
  \bibinfo {pages} {9356} (\bibinfo {year} {2012})}\BibitemShut {NoStop}%
\bibitem [{\citenamefont {Wang}\ and\ \citenamefont
  {Wang}(2011)}]{wang2011jcp}%
  \BibitemOpen
  \bibfield  {author} {\bibinfo {author} {\bibfnamefont {R.}~\bibnamefont
  {Wang}}\ and\ \bibinfo {author} {\bibfnamefont {Z.-G.}\ \bibnamefont
  {Wang}},\ }\href {https://doi.org/10.1063/1.3607969} {\bibfield  {journal}
  {\bibinfo  {journal} {J. Chem. Phys.}\ }\textbf {\bibinfo {volume} {135}}
  (\bibinfo {year} {2011})}\BibitemShut {NoStop}%
\bibitem [{\citenamefont {Sing}\ \emph {et~al.}(2014)\citenamefont {Sing},
  \citenamefont {Zwanikken},\ and\ \citenamefont {Olvera~de
  La~Cruz}}]{sing2014}%
  \BibitemOpen
  \bibfield  {author} {\bibinfo {author} {\bibfnamefont {C.~E.}\ \bibnamefont
  {Sing}}, \bibinfo {author} {\bibfnamefont {J.~W.}\ \bibnamefont
  {Zwanikken}},\ and\ \bibinfo {author} {\bibfnamefont {M.}~\bibnamefont
  {Olvera~de La~Cruz}},\ }\href {https://www.nature.com/articles/nmat4001}
  {\bibfield  {journal} {\bibinfo  {journal} {Nat. Mater.}\ }\textbf {\bibinfo
  {volume} {13}},\ \bibinfo {pages} {694} (\bibinfo {year} {2014})}\BibitemShut
  {NoStop}%
\bibitem [{\citenamefont {Hou}\ and\ \citenamefont {Qin}(2018)}]{Kevin2018}%
  \BibitemOpen
  \bibfield  {author} {\bibinfo {author} {\bibfnamefont {K.~J.}\ \bibnamefont
  {Hou}}\ and\ \bibinfo {author} {\bibfnamefont {J.}~\bibnamefont {Qin}},\
  }\href {https://doi.org/10.1021/acs.macromol.8b01616} {\bibfield  {journal}
  {\bibinfo  {journal} {Macromolecules}\ }\textbf {\bibinfo {volume} {51}},\
  \bibinfo {pages} {7463} (\bibinfo {year} {2018})}\BibitemShut {NoStop}%
\bibitem [{\citenamefont {Liu}\ \emph {et~al.}(2022)\citenamefont {Liu},
  \citenamefont {Duan},\ and\ \citenamefont {Wang}}]{Liu2022}%
  \BibitemOpen
  \bibfield  {author} {\bibinfo {author} {\bibfnamefont {L.}~\bibnamefont
  {Liu}}, \bibinfo {author} {\bibfnamefont {C.}~\bibnamefont {Duan}},\ and\
  \bibinfo {author} {\bibfnamefont {R.}~\bibnamefont {Wang}},\ }\href
  {https://www.sciencedirect.com/science/article/pii/S0032386122007996}
  {\bibfield  {journal} {\bibinfo  {journal} {Polymer}\ }\textbf {\bibinfo
  {volume} {258}},\ \bibinfo {pages} {125312} (\bibinfo {year}
  {2022})}\BibitemShut {NoStop}%
\bibitem [{\citenamefont {Grosberg}\ and\ \citenamefont
  {Kuznetsov}(1992)}]{grosberg1992}%
  \BibitemOpen
  \bibfield  {author} {\bibinfo {author} {\bibfnamefont {A.~Y.}\ \bibnamefont
  {Grosberg}}\ and\ \bibinfo {author} {\bibfnamefont {D.}~\bibnamefont
  {Kuznetsov}},\ }\href {https://pubs.acs.org/doi/abs/10.1021/ma00033a024}
  {\bibfield  {journal} {\bibinfo  {journal} {Macromolecules}\ }\textbf
  {\bibinfo {volume} {25}},\ \bibinfo {pages} {1991} (\bibinfo {year}
  {1992})}\BibitemShut {NoStop}%
\bibitem [{\citenamefont {Fredrickson}(2006)}]{fredrickson2006equilibrium}%
  \BibitemOpen
  \bibfield  {author} {\bibinfo {author} {\bibfnamefont {G.}~\bibnamefont
  {Fredrickson}},\ }\href@noop {} {\emph {\bibinfo {title} {The equilibrium
  theory of inhomogeneous polymers}}}\ (\bibinfo  {publisher} {Oxford
  University Press},\ \bibinfo {year} {2006})\BibitemShut {NoStop}%
\bibitem [{\citenamefont {Xu}\ \emph {et~al.}(2014)\citenamefont {Xu},
  \citenamefont {Ting}, \citenamefont {Kusaka},\ and\ \citenamefont
  {Wang}}]{Xu2014}%
  \BibitemOpen
  \bibfield  {author} {\bibinfo {author} {\bibfnamefont {X.}~\bibnamefont
  {Xu}}, \bibinfo {author} {\bibfnamefont {C.~L.}\ \bibnamefont {Ting}},
  \bibinfo {author} {\bibfnamefont {I.}~\bibnamefont {Kusaka}},\ and\ \bibinfo
  {author} {\bibfnamefont {Z.-G.}\ \bibnamefont {Wang}},\ }\href
  {https://doi.org/10.1146/annurev-physchem-032511-143750} {\bibfield
  {journal} {\bibinfo  {journal} {Annu. Rev. Phys. Chem.}\ }\textbf {\bibinfo
  {volume} {65}},\ \bibinfo {pages} {449} (\bibinfo {year} {2014})}\BibitemShut
  {NoStop}%
\bibitem [{\citenamefont {Bottacchiari}\ \emph {et~al.}(2022)\citenamefont
  {Bottacchiari}, \citenamefont {Gallo}, \citenamefont {Bussoletti},\ and\
  \citenamefont {Casciola}}]{Bottacchiari2022}%
  \BibitemOpen
  \bibfield  {author} {\bibinfo {author} {\bibfnamefont {M.}~\bibnamefont
  {Bottacchiari}}, \bibinfo {author} {\bibfnamefont {M.}~\bibnamefont {Gallo}},
  \bibinfo {author} {\bibfnamefont {M.}~\bibnamefont {Bussoletti}},\ and\
  \bibinfo {author} {\bibfnamefont {C.~M.}\ \bibnamefont {Casciola}},\ }\href
  {https://www.nature.com/articles/s42005-022-01055-2} {\bibfield  {journal}
  {\bibinfo  {journal} {Commun. Phys.}\ }\textbf {\bibinfo {volume} {5}},\
  \bibinfo {pages} {1} (\bibinfo {year} {2022})}\BibitemShut {NoStop}%
\bibitem [{\citenamefont {E}\ and\ \citenamefont
  {Vanden-Eijnden}(2010)}]{Weinan2010}%
  \BibitemOpen
  \bibfield  {author} {\bibinfo {author} {\bibfnamefont {W.}~\bibnamefont {E}}\
  and\ \bibinfo {author} {\bibfnamefont {E.}~\bibnamefont {Vanden-Eijnden}},\
  }\href {https://doi.org/10.1146/annurev.physchem.040808.090412} {\bibfield
  {journal} {\bibinfo  {journal} {Annu. Rev. Phys. Chem.}\ }\textbf {\bibinfo
  {volume} {61}},\ \bibinfo {pages} {391} (\bibinfo {year} {2010})}\BibitemShut
  {NoStop}%
\bibitem [{\citenamefont {Ward}\ and\ \citenamefont
  {Sweeney}(2012)}]{ward2012mechanical}%
  \BibitemOpen
  \bibfield  {author} {\bibinfo {author} {\bibfnamefont {I.~M.}\ \bibnamefont
  {Ward}}\ and\ \bibinfo {author} {\bibfnamefont {J.}~\bibnamefont {Sweeney}},\
  }\href@noop {} {\emph {\bibinfo {title} {Mechanical properties of solid
  polymers}}}\ (\bibinfo  {publisher} {John Wiley \& Sons},\ \bibinfo {year}
  {2012})\BibitemShut {NoStop}%
\bibitem [{\citenamefont {Brochard}\ \emph {et~al.}(1976)\citenamefont
  {Brochard}, \citenamefont {De~Gennes},\ and\ \citenamefont
  {Pfeuty}}]{brochard1976}%
  \BibitemOpen
  \bibfield  {author} {\bibinfo {author} {\bibfnamefont {F.}~\bibnamefont
  {Brochard}}, \bibinfo {author} {\bibfnamefont {P.}~\bibnamefont
  {De~Gennes}},\ and\ \bibinfo {author} {\bibfnamefont {P.}~\bibnamefont
  {Pfeuty}},\ }\href
  {https://jphys.journaldephysique.org/articles/jphys/abs/1976/10/jphys_1976__37_10_1099_0}
  {\bibfield  {journal} {\bibinfo  {journal} {J. Phys.}\ }\textbf {\bibinfo
  {volume} {37}},\ \bibinfo {pages} {1099} (\bibinfo {year}
  {1976})}\BibitemShut {NoStop}%
\bibitem [{\citenamefont {Arroyo}\ and\ \citenamefont
  {DeSimone}(2009)}]{Arroyo2009}%
  \BibitemOpen
  \bibfield  {author} {\bibinfo {author} {\bibfnamefont {M.}~\bibnamefont
  {Arroyo}}\ and\ \bibinfo {author} {\bibfnamefont {A.}~\bibnamefont
  {DeSimone}},\ }\href {https://link.aps.org/doi/10.1103/PhysRevE.79.031915}
  {\bibfield  {journal} {\bibinfo  {journal} {Phys. Rev. E}\ }\textbf {\bibinfo
  {volume} {79}},\ \bibinfo {pages} {031915} (\bibinfo {year}
  {2009})}\BibitemShut {NoStop}%
\bibitem [{\citenamefont {Steed}\ and\ \citenamefont
  {Gale}(2012)}]{steed2012supramolecular}%
  \BibitemOpen
  \bibfield  {author} {\bibinfo {author} {\bibfnamefont {J.~W.}\ \bibnamefont
  {Steed}}\ and\ \bibinfo {author} {\bibfnamefont {P.~A.}\ \bibnamefont
  {Gale}},\ }\href@noop {} {\emph {\bibinfo {title} {Supramolecular chemistry:
  from molecules to nanomaterials}}}\ (\bibinfo  {publisher} {John Wiley \&
  Sons},\ \bibinfo {year} {2012})\BibitemShut {NoStop}%
\bibitem [{\citenamefont {Duan}\ and\ \citenamefont {Wang}(2023)}]{Duan2023}%
  \BibitemOpen
  \bibfield  {author} {\bibinfo {author} {\bibfnamefont {C.}~\bibnamefont
  {Duan}}\ and\ \bibinfo {author} {\bibfnamefont {R.}~\bibnamefont {Wang}},\
  }\href {https://link.aps.org/doi/10.1103/PhysRevLett.130.158401} {\bibfield
  {journal} {\bibinfo  {journal} {Phys. Rev. Lett.}\ }\textbf {\bibinfo
  {volume} {130}},\ \bibinfo {pages} {158401} (\bibinfo {year}
  {2023})}\BibitemShut {NoStop}%
\end{thebibliography}%

\end{document}


\title[]
{Supplementary Material for: Kinetic Pathway and Micromechanics of Vesicle Fusion/Fission}

\author{Luofu Liu}
\affiliation{Department of Chemical and Biomolecular Engineering, University of California Berkeley, Berkeley, California 94720, United States}

\author{Chao Duan}
\affiliation{Department of Chemical and Biomolecular Engineering, University of California Berkeley, Berkeley, California 94720, United States}

\author{Rui Wang}
\email {ruiwang325@berkeley.edu}\affiliation{Department of Chemical and Biomolecular Engineering, University of California Berkeley, Berkeley, California 94720, United States}
\affiliation{Materials Sciences Division, Lawrence Berkeley National Lab, Berkeley, California 94720, United States}

\renewcommand*{\citenumfont}[1]{S#1}
\renewcommand*{\bibnumfmt}[1]{(S#1)}

\renewcommand{\theequation}{S\arabic{equation}}
\renewcommand{\thefigure}{S\arabic{figure}}

\maketitle

\section{I. Derivation of Constrained Self-consistent Field Theory for Polyelectrolyte Vesicles}

In this section, we give a detailed derivation of the key equations in the constrained SCFT for the polyelectrolyte (PE) vesicles with fixed centers-of-mass (CM). The semi-canonical partition function (as Eq. 1 in the main text) is
\begin{equation} \label{eq:partition_fucntion}
\begin{aligned}
    {\Xi} = \frac{1}{v_{\rm p}^{2N}} &\sum_{n_{\gamma}=0}^\infty \prod_{\gamma} \frac{e^{\mu_{\gamma} n_{\gamma}}}{n_{\gamma}!v_{\gamma}^{n_{\gamma}}}\prod_{j=1}^2\int {\rm \hat{D}}\{\mathbf{R}_j\} \prod_{\kappa = 1}^{n_{\gamma}} \int{\rm d}\mathbf{r}_{\gamma,\kappa} \exp(-{\beta} \mathcal{H})\prod_{\mathbf{r}}\delta \left[\hat{\phi}_{\rm p}(\mathbf{r})+\hat{\phi}_{\rm s}(\mathbf{r})-1 \right]  \\
    &\times \delta\left[\frac{1}{N}\int_{0}^{N}{\rm d}s\mathbf{R}_1(s)-\boldsymbol{\xi}_1\right]\delta\left[\frac{1}{N}\int_{0}^{N}{\rm d}s\mathbf{R}_2(s)-\boldsymbol{\xi}_2\right] \\
\end{aligned}
\end{equation}
where $\gamma = {\rm s}, \pm$ stands for small molecules, i.e., solvents, cations and anions. $\int \hat{D}\{\mathbf{R}_j\}$ denotes the integration over the Gaussian-weighted chain configurations, $\int{\rm d}\mathbf{r}_{\gamma,\kappa}$ denotes the integration over the degree of freedom of small molecules. $v_{\rm p}$ and $v_\gamma$ are the volumes of the chain segments and small molecules, respectively. For simplicity, we assume $v_{\rm p} = v_{\rm s} = v$. $\hat{\phi}_{\rm p} = \hat{\phi}_{\rm p1} + \hat{\phi}_{\rm p2} = v_{\rm p}\sum_{j = 1}^2 \int_0^N {\rm d}s \delta(\mathbf{r} - \mathbf{R}_j(s)) $ is the total instantaneous volume fractions of PE. $\hat{\phi}_{\rm s} = v_{\rm s}\sum_{\kappa = 1}^{n_{\rm s}}\delta(\mathbf{r} - \mathbf{r}_{{\rm s},\kappa})$ is the instantaneous volume fraction of solvent. The $\delta$ functional in the first line of Eq. \ref{eq:partition_fucntion} accounts for the incompressibility. The Hamiltonian $\mathcal{H}$ is given by
\begin{equation} \label{eq:Hamiltonian}
{\beta}\mathcal{H} =  \frac{\chi}{v}\int{\rm d}{\mathbf{r}}\hat{\phi}_{\rm p}(\mathbf{r})\hat{\phi}_{\rm s}(\mathbf{r}) + \frac{1}{2}\int {\rm d}{\mathbf{r}}\int {\rm d}{\mathbf{r^{'}}} \hat{\rho}_{\rm c}(\mathbf{r}) C(\mathbf{r}, \mathbf{r^{'}})\hat{\rho}_{\rm c}(\mathbf{r}^{'})
\end{equation}
Here, the first term on the r.h.s of Eq. \ref{eq:Hamiltonian} comes from the short-ranged polymer-solvent interaction characterized by a Flory-Huggins $\chi$ parameter, and the second term comes from the electrostatic interactions between charged species. $\hat{\rho}_{\rm c}(\mathbf{r}) = z_{+}\hat{c}_{+}(\mathbf{r}) - z_{-}\hat{c}_{-}(\mathbf{r}) - \alpha \hat{\phi}_{\rm_p}(\mathbf{r})/v$ is the instantaneous local charge density, with $\hat{c}_{\pm}(\mathbf{r})$ the instantaneous concentration of cations and anions. $C(\mathbf{r}, \mathbf{r^{'}})$ is the Coulomb operator, satisfying
\begin{equation}
-\nabla \cdot[\epsilon(\mathbf{r})\nabla C(\mathbf{r},\mathbf{r^{'}})] = \delta(\mathbf{r} -\mathbf{r^{'}})
\end{equation}
where $\epsilon(\mathbf{r}) = \epsilon_0\epsilon_{\rm r}(\mathbf{r})/\beta e^2$ is the rescaled permittivity, where $\epsilon_0$ is the vacuum permitivity, $e$ is the elementary charge and $\epsilon_{\rm r}(\mathbf{r})$ is local dielectric constant. The two $\delta$ functions in the second line of the partition function, $\delta [N^{-1}\int_0^N{\rm d}s\mathbf{R}_j(s)-\boldsymbol{\xi}_j]$, are introduced to enforce the center-of-mass of the Vesicle $j$ at the position $\xi_j$ \cite{Liu2022, grosberg1992}.

Here, we perform the standard self-consistent field approach  \cite{fredrickson2006equilibrium}, which involves (1) decoupling the interacting system into noninteracting chains in the fluctuating fields by identity transformation; and (2) replacing the functional integration over the fluctuating fields by the saddle point approximation. To perform the identity transformation, firstly, the following identities will be inserted into the partition function:
\begin{equation}
\begin{aligned}
        1 &\equiv \int {\rm D}\phi_{\sigma} \prod_{\mathbf{r}}\delta \left[\phi_\sigma(\mathbf{r}) - \hat{\phi}_{\sigma}(\mathbf{r}) \right]\\
    &= \int {\rm D}\phi_\sigma {\rm D} w_\sigma \exp \left\{\frac{1}{v}\int{\rm d}{\mathbf{r}}iw_\sigma(\mathbf{r}) \cdot \left[\phi_\sigma(\mathbf{r})-\hat{\phi}_\sigma(\mathbf{r}) \right] \right\},\ \sigma = {\rm p}1,\ {\rm p}2,\ \rm{s}
\end{aligned}
\end{equation}
where the Fourier transform of the $\delta$ functionals introduces $w_\sigma(\mathbf{r})$ as the conjugate potential field of the density field $\phi_\sigma(\mathbf{r})$ for the species $\sigma$. Then, similarly, Fourier transform is performed on the $\delta$ functional in $\Xi$, and introduces the potential field $\eta(\mathbf{r})$ to enforce incompressibility and uniform force field $\mathbf{f}_j$ to maintain the fixed CM of the Vesicle $j$:
\begin{subequations}
    \begin{align}
        \prod_{\mathbf{r}} \delta\left[\hat{\phi}_{\rm p}(\mathbf{r})+\hat{\phi}_{\rm s}(\mathbf{r})-1\right] &= \int {\rm D}\eta \exp\left\{\frac{1}{v}\int {\rm d} \mathbf{r} i\eta(\mathbf{r}) \left[\hat{\phi}_{\rm p}(\mathbf{r}) + \hat{\phi}_{\rm s}(\mathbf{r}) -1 \right] \right\} \\
        \delta \left[\frac{1}{N} \int_0^{N} {\rm d}s \mathbf{R}_j(s) - \boldsymbol{\xi}_j \right] &=\int {\rm d} \mathbf{f}_j \exp \left\{i\mathbf{f}_j \cdot \left[ \frac{1}{N} \int_0^{N} {\rm d}s\mathbf{R}_j(s) - \boldsymbol{\xi}_j \right]   \right\}
    \end{align}
\end{subequations}
Moreover, to decouple the electrostatic interactions between charged particles, we perform the Hubbard-Stratonovich transform on the electrostatic part of the Hamiltonian:
\begin{equation}\label{eq:HS_transform}
\begin{aligned}
    &\exp\left\{-\frac{1}{2}\int {\rm d} \mathbf{r} \int {\rm d} \mathbf{r^{'}}  \hat{\rho}_{\rm c} (\mathbf{r}) C(\mathbf{r},\mathbf{r^{'}}) \hat{\rho}_{\rm c} (\mathbf{r^{'}}) \right\} = \\
    & \mathcal{N}_{\psi} \int {\rm D} \psi \exp \left\{ -\frac{1}{2} \int {\rm d}\mathbf{r} \int {\rm d} \mathbf{r^{'}} \psi(\mathbf{r})C^{-1}(\mathbf{r},\mathbf{r^{'}})\psi(\mathbf{r^{'}}) -i \int {\rm d}\mathbf{r} \hat{\rho}_{\rm c}(\mathbf{r}) \psi(\mathbf{r}) \right\}
\end{aligned}
\end{equation}
where $\mathcal{N}_\psi^{-1} = \int D \psi \exp \{-(1/2) \int {\rm d}\mathbf{r} \int {\rm d} \mathbf{r^{'}} \psi(\mathbf{r}) C^{-1}(\mathbf{r},\mathbf{r^{'}}) \psi(\mathbf{r^{'}})\}$ is the normalization constant. $C^{-1}(\mathbf{r},\mathbf{r^{'}}) = - \nabla \cdot [\epsilon(\mathbf{r})\nabla]\delta(\mathbf{r}-\mathbf{r^{'}})$ is the inverse of Coulomb operator $C(\mathbf{r},\mathbf{r^{'}})$, and $\psi$ is the dimensionless eletrostatic field rescaled by $k_{\rm B}T/e$. Further simplification of Eq. \ref{eq:HS_transform} gives
\begin{equation}
\begin{aligned}
    &\exp\{-\frac{1}{2}\int {\rm d} \mathbf{r} \int {\rm d} \mathbf{r^{'}}  \hat{\rho}_{\rm c} (\mathbf{r}) C(\mathbf{r},\mathbf{r^{'}}) \hat{\rho}_{\rm c} (\mathbf{r^{'}}) \} = \\
    & \mathcal{N}_{\psi} \int {\rm D} \psi \exp \left\{ \int {\rm d}\mathbf{r} \left[\frac{1}{2}\epsilon(\mathbf{r})|\nabla \psi(\mathbf{r})|^2 - \left( z_{+} \hat{c}_{+}(\mathbf{r}) - z_{-}\hat{c}_{-}(\mathbf{r}) - \frac{\alpha}{v}\hat{\phi}_{\rm p}(\mathbf{r}) \right)i\psi(\mathbf{r})  \right] \right\}
\end{aligned}
\end{equation}

By performing the identity transformations, now we obtain the field-based partition function as
\begin{equation} \label{eq:complete_PF}
    \begin{aligned}
        \Xi &= \int {\rm D}\phi_{{\rm p}1}  {\rm D}w_{{\rm p}1}  {\rm D}\phi_{{\rm p}2}  {\rm D}w_{{\rm p}2} {\rm D}\phi_{\rm s} {\rm D}w_{\rm s} {\rm D}\eta {\rm D}\psi {\rm d}\mathbf{f}_1 {\rm d}\mathbf{f}_2
        \frac{1}{v^{2N}}\sum_{n_\gamma=0}^\infty \prod_{\gamma} \frac{e^{\mu_{\gamma} n_{\gamma}}}{n_{\gamma}!v^{n_{\gamma}}}\int \prod_{j=1}^2 {\rm D}\{\mathbf{R}_j\} \prod_{\kappa = 1}^{n_{\gamma}} {\rm d}\mathbf{r}_{\gamma,\kappa}  \\
        & \exp \left\{ - \sum_{j=1}^2 \frac{3}{2b^2}\int_0^N {\rm d}s \left(\frac{\partial \mathbf{R}_j(s)}{\partial s} \right)^2 \right\} \\
        &\exp\left\{ \frac{1}{v}\int{\rm d}{\mathbf{r}} \left[-\chi \phi_{\rm p}(\mathbf{r})\phi_{\rm s}(\mathbf{r}) + \sum_{j = 1}^2iw_{{\rm p}j}(\mathbf{r})(\phi_{{\rm p}j}(\mathbf{r})-\hat{\phi}_{{\rm p}j}(\mathbf{r})) + i w_{\rm s}(\mathbf{r})(\phi_{\rm s}(\mathbf{r})-\hat{\phi}_{\rm s}(\mathbf{r})) \right] \right\} \\
        &\exp\left\{ \int {\rm d}\mathbf{r} \left[-\frac{1}{2}\epsilon(\mathbf{r})|\nabla \psi(\mathbf{r})|^2 - \left( z_{+} \hat{c}_{+}(\mathbf{r}) - z_{-}\hat{c}_{-}(\mathbf{r}) - \frac{\alpha}{v}{\phi}_{\rm p}(\mathbf{r}) \right)i\psi(\mathbf{r})  \right] \right\} \\
        &\exp\left\{\frac{1}{v}\int {\rm d}\mathbf{r}i \eta(\mathbf{r}) \left[\phi_{\rm p}(\mathbf{r}) + \phi_{\rm s}(\mathbf{r}) -1 \right]  \right\} \exp \left\{\sum_{j=1}^2 i\mathbf{f}_j \cdot \left[ \frac{1}{N} \int_0^{N} {\rm d}s\mathbf{R}_j(s) - \boldsymbol{\xi}_j \right] \right\}
    \end{aligned}
\end{equation}
where $\phi_{\rm p}(\mathbf{r}) \equiv \phi_{{\rm p}1}(\mathbf{r})+\phi_{{\rm p}2}(\mathbf{r})$ is the total polymer volume fraction. Note that here we have expanded the Gaussian-weighted chain integral as
\begin{equation}
\int \hat{\rm D}\{ \mathbf{R}_j \} = \int {\rm D}\{ \mathbf{R}_j \} \exp \left\{ -\frac{3}{2b^2}\int_0^N {\rm d}s \left(\frac{\partial \mathbf{R}_j(s)}{\partial s} \right)^2 \right\}
\end{equation}

We can further simplify and re-arrange the terms and write Eq. \ref{eq:complete_PF} in a more compact form:
\begin{equation}
    \begin{aligned}
        \Xi&= \int {\rm D}\phi_{{\rm p}1}  {\rm D}w_{{\rm p}1}  {\rm D}\phi_{{\rm p}2}  {\rm D}w_{{\rm p}2} {\rm D}\phi_{\rm s} {\rm D}w_{\rm s} {\rm D}\eta {\rm D}\psi {\rm d}\mathbf{f}_1 {\rm d}\mathbf{f}_2 \
        Q_{{\rm p}1}Q_{{\rm p}2}\exp(e^{\mu_{\rm s}}Q_{\rm s}) \\ &\exp\left\{\frac{1}{v}\int {\rm d}\mathbf{r}\left[-\chi \phi_{\rm p}(\mathbf{r}) \phi_{\rm s}(\mathbf{r}) + iw_{{\rm p}1}(\mathbf{r})\phi_{{\rm p}1}(\mathbf{r})+iw_{{\rm p}2}(\mathbf{r})\phi_{{\rm p}2}(\mathbf{r})+ iw_{\rm s}(\mathbf{r})\phi_{\rm s}(\mathbf{r}) \right]  \right\} \\
        &\exp \left\{ \int {\rm d}\mathbf{r} \left[\lambda_{+}e^{-z_{+}i\psi(\mathbf{r})} +  \lambda_{-}e^{z_{-}i\psi(\mathbf{r})} - \frac{1}{2}\epsilon(\mathbf{r})|\nabla \psi(\mathbf{r})|^2 + \frac{\alpha}{v} \phi_{\rm p}(\mathbf{r})i\psi(\mathbf{r}) \right]  \right\}  \\ 
        &\exp\left\{\frac{1}{v}\int {\rm d}\mathbf{r}i \eta(\mathbf{r}) \left[\phi_{\rm p}(\mathbf{r}) + \phi_{\rm s}(\mathbf{r}) -1 \right]  \right\}
    \end{aligned}
\end{equation}
where $Q_{\rm s}$ is the solvent partition function, given by
\begin{equation}\label{solvent_PF}
    Q_{\rm s} = \frac{1}{v}\int {\rm d}\mathbf{r}\exp\left[-iw_{\rm s}(\mathbf{r}) \right]
\end{equation}
and $Q_{{\rm p}j}$ is the single-chain partition function for polymer $j$, given by
\begin{equation}\label{eq:single_chain_PF}
    Q_{{\rm p}j} = \frac{1}{v^{N}}\int \hat{\rm D} \{\mathbf{R}_j\}
    \exp \left\{\int_0^{N}{\rm d}s \left[ - iw_{{\rm p}j}\left(\mathbf{R}_j(s)\right)+i\frac{\mathbf{f}_j}{N}\cdot(\mathbf{R}_j(s) - \boldsymbol{\xi}_j) \right]  \right \}
\end{equation}
Eq. \ref{eq:single_chain_PF} can be re-expressed in terms of chain propagator $q_{j}(\mathbf{r},s)$ as
\begin{equation}
    Q_{{\rm p},j} = \frac{1}{v} \int {\rm d} \mathbf{r} q_{j} (\mathbf{r},N)
\end{equation}
where $q_{j}(\mathbf{r},s)$ satisfies the modified diffusion equation
\begin{equation}\label{eq:propagator}
    \left\{ \frac{\partial }{\partial s} - \frac{b^2}{6}\nabla^2 + i\left[ w_{{\rm p}j}(\mathbf{r})-\frac{\mathbf{f}_j}{N}\cdot(\mathbf{r}-\boldsymbol{\xi}_j) \right]  \right\}q_j(\mathbf{r},s) = 0
\end{equation}

The couplings between fields make the direct evaluation of Eq. \ref{eq:complete_PF} intractable, so we approximate $\Xi$ with maximum of the integrand in the functional integration. The approximation is valid when thermal fluctuation is small compared to the maximum which is the case for polymer globules in poor solvents. And the corresponding fields are found by functional minimization of the field-based Hamiltonian via saddle-point approximation which gives the following SCF equations:
\begin{subequations}\label{scfs_SI}
\begin{align}
&w_{{\rm p}j}(\mathbf{r}) = \chi \phi_{\rm s}(\mathbf{r}) - \eta(\mathbf{r}) - v\frac{\partial \epsilon}{\partial \phi_{\rm p}} \frac{|\nabla \psi(\mathbf{r})|^2}{2} - \alpha \psi(\mathbf{r}) \label{scfSI_a}\\
&w_{\rm s}(\mathbf{r}) = \chi \phi_{\rm p}(\mathbf{r}) - \eta(\mathbf{r}) \label{scfSI_b}\\
&\phi_{{\rm p}j}(\mathbf{r}) = \frac{1}{Q_{{\rm p}j}}\int_0^{N}{\rm d}s q_j(\mathbf{r},s)q_j(\mathbf{r},N-s) \ j=1,\ 2\label{scfSI_c} \\
&\phi_{\rm s}(\mathbf{r}) = \exp\left[\mu_{\rm s}-w_{\rm s}(\mathbf{r})\right] \label{scfSI_d} \\
&\phi_{\rm p}(\mathbf{r}) + \phi_{\rm s}(\mathbf{r}) = 1 \label{scfSI_e}\\
& -\nabla \cdot [\epsilon(\mathbf{r})\nabla \psi(\mathbf{r})] = z_{+}\lambda_{+}e^{-z_{+}\psi(\mathbf{r})} - z_{-}\lambda_{-}e^{z_{-}\psi(\mathbf{r})} - \frac{\alpha}{v} \phi_{\rm p}(\mathbf{r}) \\
&{\int {\rm d}\mathbf{r}\phi_{{\rm p}j}(\mathbf{r})\cdot(\mathbf{r}-\boldsymbol{\xi}_j)} = \mathbf{0}, \ j=1,\ 2 \label{scfSI_g}
\end{align}
\end{subequations}
Equations \ref{scfs_SI} becomes Eqs. 3 in the main text when we replace $w_{{\rm p}j}(\mathbf{r})$ with $w_{\rm p}(\mathbf{r})$ and make use of the incompressibility condition. Note that we have turned the conjugate fields into real values by multiplying $-i$ to them considering the fact that the saddle point of the original fields are purely imaginary. The semi-canonical free energy is given by
\begin{equation}\label{eq:free_energy_SI}
    \begin{aligned}
     F &= -e^{\beta\mu_{\rm s}}Q_{\rm s} -  \ln Q_{{\rm p}1}  -  \ln Q_{{\rm p}2}\\
    &+ \frac{1}{v}\int {\rm d}{\mathbf{r}}\left\{\chi\phi_{\rm p}(\mathbf{r})\phi_{\rm s}(\mathbf{r})-w_{\rm p}(\mathbf{r})\phi_{\rm p}(\mathbf{r})-w_{\rm s}(\mathbf{r})\phi_{\rm s}(\mathbf{r}) - \eta(\mathbf{r})\left[\phi_{\rm p}(\mathbf{r})+\phi_{\rm s}(\mathbf{r})-1\right] \right\} \\
    & - \int {\rm d}\mathbf{r}\left\{ \lambda_{+} e^{-z_{+}\psi(\mathbf{r})} + \lambda_{-} e^{z_{-}\psi(\mathbf{r})} + \frac{\epsilon(\mathbf{r})}{2}|\nabla \psi(\mathbf{r})|^2 + \frac{\alpha}{v}\phi_{\rm p}(\mathbf{r})\psi(\mathbf{r}) \right\} \\
    \end{aligned}
\end{equation}
Similarly, Eq. \ref{eq:free_energy_SI} can be further expressed as Eq. 5 in the main text using the incompressibility condition.

\section{II. Numerical Details for Solving Self-consistent Equations}

Based on the symmetry of the possible configurations, we use the cylindrical coordinate in the numerical calculation. The modified diffusion equation (Eq. 4 in the main text) is solved by the approximate-factorization implicit (AFI) method, whereas the Poisson-Boltzmann equation (Eq. 3d in the main text) is solved by the alternating-direction implicit (ADI) method. The number of points that the chain contour has been discretized is set to be $N_{\rm s} = 500$. The grid lattices are set to be $\Delta r = \Delta z = b/3$ to guarantee sufficient spatial resolution.

The equilibrium structure, free energy, and force can be obtained by solving Eqs. 3 and Eq. 4 iteratively until convergence. $\mu_{\rm s}$ is set to be -1 such that the free energy of the reservoir of pure salt solution is 0. The initial seed for the parent vesicle at $L = 0$ is a spherical globule, which finally evolves to the vesicle structure as iteration proceeds. We select the converged structure for the previous $L$ to be the initial seed for the next $L$ when computing the free energy curves. The interval of $L/b$ is 1.0. The density profiles for two separated child vesicles are obtained by solving the system of one vesicle while applying reflection boundary condition on a boundary of the box on the $z$ direction. We use the following strategy to update the fields. Fields conjugate to the density of PE and solvent molecules are updated by a simple mixing rule, i. e., $w_{\rm p,s}^{new} \leftarrow \lambda w_{\rm p,s}^{new} + (1-\lambda) w_{\rm p,s}^{old}$. The field conjugate to the incompressibility condition is updated by $\eta^{new} \leftarrow \eta^{\rm old} + \kappa [\phi_{\rm p} + \phi_{\rm s} - 1]$, where the second term on the r.h.s is adopted to reinforce the incompressibility.
$\lambda = 0.01$ and $\kappa = 1.0$ are chosen the our calculation. The relative errors for the free energy and incompressibility condition are set to be below $10^{-10}$ and $10^{-5}$, respectively. 

\section{III. Relating Surface Energy to Neck Length}

\begin{figure}[h] 
\centering
\includegraphics[width=0.5\textwidth]{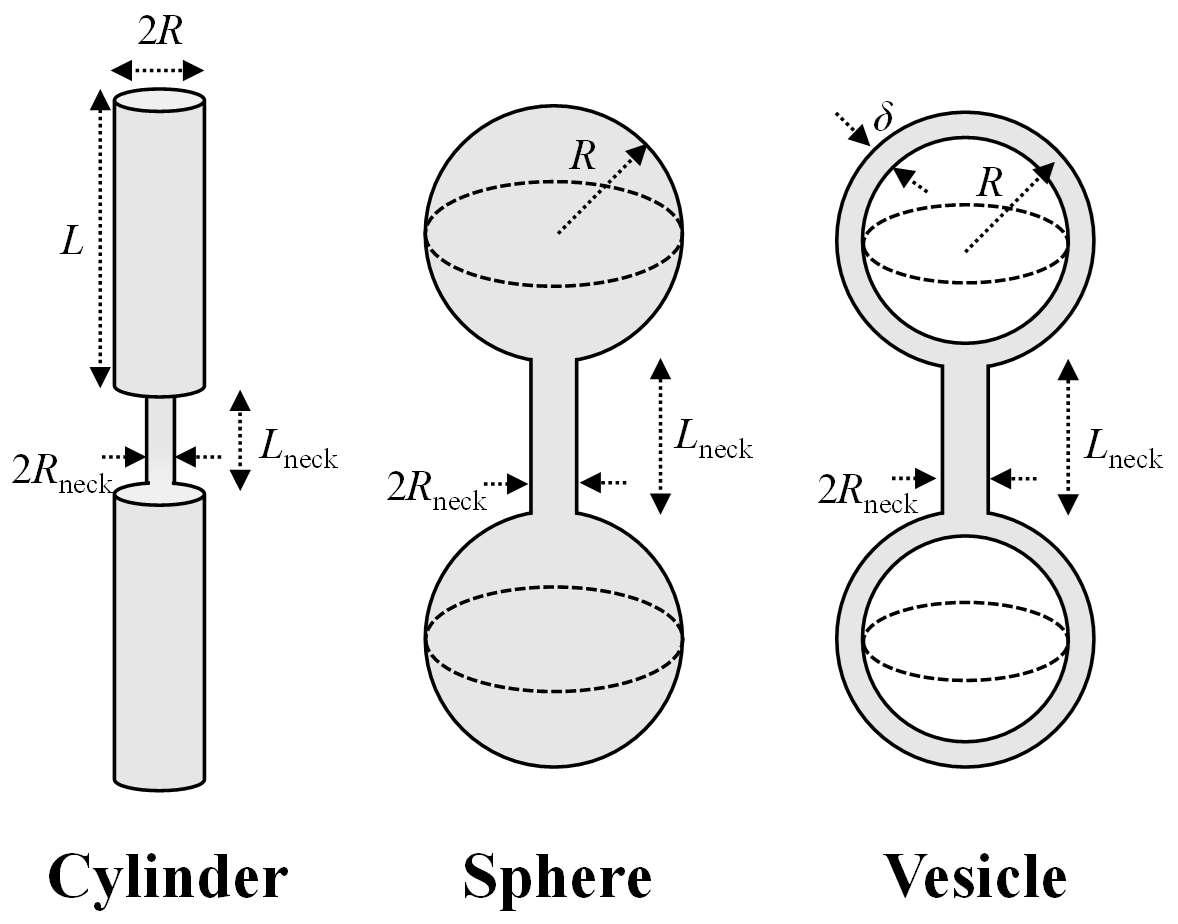}
\caption{The necking structure of separating cylinders, spheres, and vesicles approximated by standard geometries. For each structure, the neck is modeled by a column of radius $R_{\rm neck}$ and length $L_{\rm neck}$, and the rest parts are modeled by columns (radius $R$ and length $L$), spheres (radius $R$), and spherical shells (radius $R$ and thickness $\delta$) respectively.}
\label{figure:appendix_geometry}
\end{figure}

Here, we relate the surface energy to neck length to explain different necking behaviors when separating two cylinders, spheres, and vesicles. We first assume a neck in each structure and then analyze the dependence of surface energy $U_{\rm surf}$ on the neck length. The neck is modeled as a thin column with a length $L_{\rm neck}$ and radius $R_{\rm neck}$, and the rest parts of the structures are approximated by standard geometries, as shown in Fig. \ref{figure:appendix_geometry}.

The surface energy of each structure is proportional to the surface area:
\begin{equation}\label{eq:U_surf}
U_{\rm surf} \propto \left\{
\begin{aligned}
4 \pi RL + 2\pi R_{\rm neck}L_{\rm neck} & , & {\rm Cylinder} \\
8\pi R^2  + 2\pi  R_{\rm neck}L_{\rm neck}  & , & {\rm Sphere} \\
16\pi R^2  + 2\pi  R_{\rm neck}L_{\rm neck}  & , & {\rm Vesicle}
\end{aligned}
\right.
\end{equation}
It should be noted that (1) the contribution from the base faces of the columns is neglected in the cylinder case due to high aspect ratio, and (2) the inner and outer surface areas of the cavities in the vesicle case are approximated to the the same under the assumption $\delta \ll R$. Assuming the total PE volume fraction $\phi_{\rm p}$ is uniformly distributed in the structures, the number of monomers in the neck $n_{\rm neck}$ and the rest of the structures $n-n_{\rm neck}$ can be written as follows according to geometric relations:
\begin{equation}\label{eq:neck_monomers}
    n_{\rm neck} = \pi R_{\rm neck}^2 L_{\rm neck}\phi_{\rm p} / v
\end{equation}
and
\begin{equation}\label{eq:rest_monomers}
n - n_{\rm neck} = \left\{
\begin{aligned}
2\pi R^2 L \phi_{\rm p}/v & , & {\rm Cylinder} \\
\frac{8}{3}\pi R^3 \phi_{\rm p}/v  & , & {\rm Sphere} \\
8\pi R^2 \delta \phi_{\rm p} /v  & , & {\rm Vesicle}
\end{aligned}
\right.
\end{equation}
where $n$ is the total number of monomers and $v$ is the monomer volume. From the Eqs. \ref{eq:U_surf} to \ref{eq:rest_monomers}, it is straightforward to show
\begin{equation}
U_{\rm surf} \propto \left\{
\begin{aligned}
&\frac{2v}{\phi_{\rm p}} \left[\frac{n}{R} + n_{\rm neck}(\frac{1}{R_{\rm neck}} - \frac{1}{R})\right]   , & {\rm Cylinder} \\
&(8\pi)^{1/3}(\frac{3v}{\phi_{\rm p}})^{2/3}(n-n_{\rm neck})^{2/3} + \frac{2vn_{\rm neck}}{\phi R_{\rm neck}}   , & {\rm Sphere} \\
&\frac{2v}{\phi_{\rm p}} \left[\frac{n}{\delta} + n_{\rm neck}(\frac{1}{R_{\rm neck}} - \frac{1}{\delta})\right]  , & {\rm Vesicle}
\end{aligned}
\right.
\end{equation}
and the following differential relations
\begin{equation}
\frac{\partial U_{\rm surf}}{\partial L_{\rm neck}} \propto \frac{\partial U_{\rm surf}}{\partial n_{\rm neck}}  \propto \left\{
\begin{aligned}
\frac{1}{R_{\rm neck}} - \frac{1}{R} & , & {\rm Cylinder} \\
\frac{1}{R_{\rm neck}} - \frac{1}{R}  & , & {\rm Sphere} \\
\frac{1}{R_{\rm neck}} - \frac{1}{\delta}  & , & {\rm Vesicle}
\end{aligned}
\right.
\end{equation}
For separating two cylinders or spheres, $\frac{\partial U_{\rm surf}}{\partial L_{\rm neck}}$ is positive since $R \gg R_{\rm neck}$, indicating a large surface energy penalty to grow a neck in the middle of the structure. However, $\frac{\partial U_{\rm surf}}{\partial L_{\rm neck}}\approx 0$ for the case of vesicle due to the fact that $R_{\rm neck} \approx \delta$. Different necking behaviors when separating cylinders, spheres and vesicles are essentially determined by their different geometric features.
